\documentclass[%
reprint, 
pre,
twocolumn
,showpacs
,preprintnumbers,
amsmath,
amssymb,
floatfix
]{revtex4-1}
\usepackage{mciteplus}

\usepackage[dvips]{graphicx}    
\usepackage[normalem]{ulem}       
\usepackage{natbib}
\usepackage[usenames]{color} 
\usepackage{multirow} 
\usepackage{array}

\newcommand{\be}{\begin{equation}}      
\newcommand{\bea}{\begin{eqnarray}}
\newcommand{\ee}{\end{equation}}
\newcommand{\eea}{\end{eqnarray}}
\newcommand{\bdm}{\begin{displaymath}}
\newcommand{\edm}{\end{displaymath}}
\newcommand{\half}{{\textstyle \frac{1}{2}}}    
\newcommand{\third}{{\textstyle{\frac{1}{3}}}}
\newcommand{\ten}[1]{\mathbf{#1}{}}     

\newcommand{\ellbold}{\mbox{\boldmath{$\ell$}}}
\newcommand{\lambdabold}{\mbox{\boldmath{$\lambda$}}}

\renewcommand{\vec}[1]{{\mathbf #1}}        
\newcommand{\Id}{\mathbf{I}}

\begin{document}

\title{A numerical study of stretched smectic-$A$ elastomer sheets}

\author{A. W. Brown}
\affiliation{SEPnet and the Department of Physics, Faculty of
  Engineering and Physical Sciences, University of Surrey, Guildford,
  GU2 7XH, U.K.}

\author{J. M. Adams} 
\affiliation{SEPnet and the Department of Physics, Faculty of
  Engineering and Physical Sciences, University of Surrey, Guildford,
  GU2 7XH, U.K.}

\begin{abstract}

  We present a numerical study of stretching monodomain smectic-$A$
  elastomer sheets, computed using the finite element method. When
  stretched parallel to the layer normal the microscopic layers in
  smectic elastomers are unstable to a transition to a buckled
  state. We account for the layer buckling by replacing the
  microscopic energy with a coarse grained effective free energy that
  accounts for the fine scale deformation of the layers. We augment
  this model with a term to describe the energy of deforming buckled
  layers, which is necessary to reproduce the experimentally observed
  Poisson's ratios post-buckling. We examine the spatial distribution
  of the microstructure phases for various stretching angles relative
  to the layer normal, and for different length-to-width aspect
  ratios. When stretching parallel to the layer normal the majority of
  the sample forms a bi-directionally buckled microstructure, except
  at the clamps where uni-directional microstructure is
  predicted. When stretching at small inclinations to the layer normal
  the phase of the sample is sensitive to the aspect ratio of the
  sample, with the bi-directionally buckled phase persistent to large
  angles only for small aspect ratios. We relate these theoretical
  results to experiments on smectic-$A$ elastomers.

\end{abstract}
\pacs{83.80.Va, 61.30.Vx, 46.32.+x, 02.70.Dh}
\maketitle


\section{Introduction}

\begin{figure}[!htb]
\begin{center}
  \includegraphics[width = 0.3\textwidth]{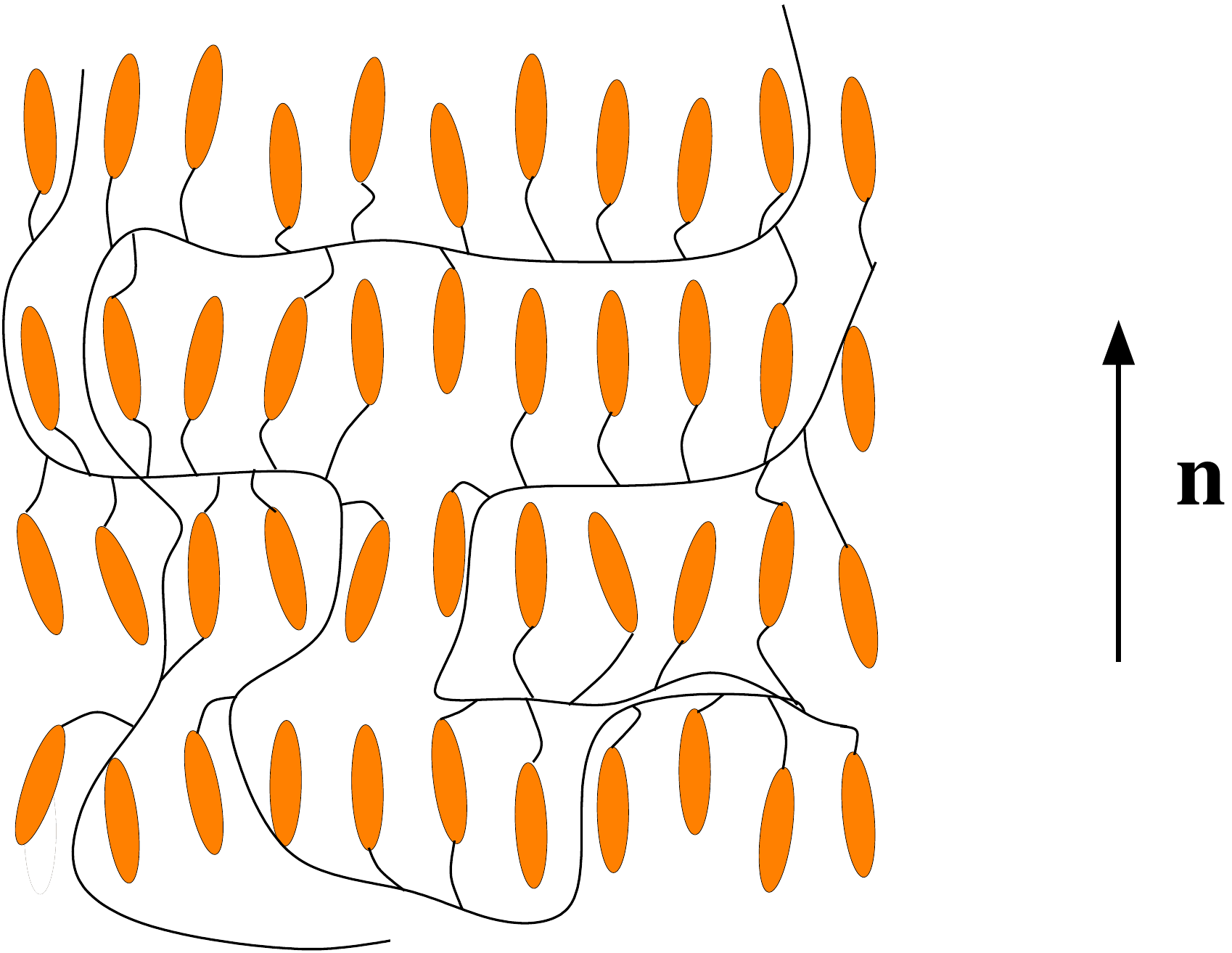}
\end{center}
\caption{An illustration of the Sm-$A$ phase in a side chain LCE. The
  layer normal $\mathbf{n}$ is parallel to the molecular
  orientation. }
\label{fig:smlayers}
\end{figure}

Liquid crystal elastomers (LCEs) are rubbery materials that are
composed of liquid crystalline polymers (LCPs) crosslinked into a
network. The rod-like mesogens incorporated into the LCPs have random
orientations in the high temperature isotropic phase, but can adopt
the canonical liquid crystalline phases at lower temperatures. The
liquid crystal phase of the mesogens plays a crucial role in the
mechanical properties of the LCE. We will focus here on the
smectic-$A$ (Sm-$A$) phase, where the mesogens form a layered
structure with the layer normal parallel to the molecular orientation
as shown in Fig.~\ref{fig:smlayers}.

In the absence of the polymer network, the layers of the liquid Sm-$A$
phase are unstable to a buckling instability when strained parallel to
their layer normal \cite{clarkmeyer1973}. Models of layered materials,
containing free energy penalties for layer curvature and layer
dilation, exhibit layer buckling \cite{singer2000}. The models predict
that just after the buckling threshold strain the layer modulation in
a single direction is degenerate with bi-directional modulation. At
larger strain this degeneracy is removed, and bi-directional modulation
is lower in energy. This theory of bi-directional buckling is
consistent with experiments on liquid smectics, where two directions
of buckling are observed in X-ray scattering patterns
\cite{ribotta1977}. The layer buckling is relaxed away in liquid
smectics by the propagation of dislocations into the layers that
relieve the strain. The smectic layer modulus in liquids is typically
$B \approx 10^7 $Pa \cite{ribotta1977}.

The first single domain Sm-$A$ elastomers was based on side chain
liquid crystalline polymers (see Fig.~\ref{fig:smlayers}) and reported
by Nishikawa \textit{et al.}  \cite{Nishikawa:97b}. A single domain
was obtained by crosslinking the elastomer subjected to a uniaxial
mechanical stress, which serves to align the layers, and hence form an
optically transparent elastomer. The sample remains transparent on
stretching perpendicular to the layer normal, and has Poisson's ratios
of $(0,1)$, indicating that the number of layers is preserved and that
the deformation is accommodated within the layers. The modulus is of
order $\mu \sim 10^5$Pa for this deformation. The elastomer is
initially much stiffer when stretched parallel to the layer normal,
having the same modulus, $B$, as the liquid smectic, and Poisson's
ratios of $(\half, \half)$.
On stretching parallel to the layer normal their elastic modulus drops
sharply to $\sim \mu$ above a threshold strain of a few percent, where
the elastomer becomes cloudy \cite{nishikawa1999}. The X-ray
scattering pattern indicates that the layers are buckled, and the
reduction in X-ray intensity shows that the modulation is in more than
one direction as layers tilt out of the X-ray scattering plane. Unlike
the liquid smectic phase the layer buckling cannot be relaxed away by
the propagation of defects. More detailed X-ray studies of similar
side chain systems reveal that the layers behave as if they are
embedded in the rubber matrix \cite{PhysRevE.82.031705}. Later
experiments on side chain systems with different chemistry have shown
the same threshold behaviour but the samples remain transparent
throughout the deformation \cite{MARC:MARC200600640}. The behaviour at
the threshold has been shown to depend on the type of smectic ordering
present \cite{MARC:MARC200700557}, and on the degree of crosslinking
in the elastomer \cite{PhysRevE.83.041703}. Some side chain systems
show behaviour closer to isotropic rubbers
\cite{doi:10.1021/ma000841q,PhysRevE.65.041707}, thought to be caused
by the interpenetration of smectic layers \cite{stannarius2006}. Here
we will focus on the Nishikawa type samples.

Smectic elastomers with a main chain polymer architecture (where the
mesogens are incorporated directly into the backbone) have
contrasting behaviour to side chain systems. The difference between
the elastic moduli in the parallel and perpendicular directions is not
as great, and the X-ray scattering patterns show increased ordering on
stretching parallel to the layer normal \cite{MARC:MARC200700210}. It
is thought that hairpin defects -- sharp reversals in the chain
orientation -- play a crucial role in the softening behaviour rather
than layer buckling \cite{ishige2008, MARC:MARC200700210,
  adams2005}. The smectic layers do not seem to be strongly coupled to
the rubber matrix, and hence do not behave as embedded
planes. Experiments suggest that smectic layers in smectic-$C$ main
chain elastomers are also only weakly coupled to the rubber matrix
\cite{sanchezferrer2011}.

Theoretical models of smectic elastomers have been successful in
describing the mechanical behaviour of side chain systems.
Phenomenological models developed using Lagrangian elasticity theory
\cite{stenull:011706}, and statistical physics \cite{adams:05} both
describe the buckling behaviour of smectic elastomers. These theories
are equivalent for small strains once the strain induced tilt of the
director is included
\cite{kramer:021704,stenull:021705,PhysRevE.78.011703}. The buckling
instability predicted by these models is symptomatic of a non-convex
free energy function. The free energy is minimized by a fine scale
mixture of deformations whose average is the macroscopic
deformation. We will not resolve this microstructure but will use a
coarse grained Sm-$A$ free energy based on the local deformation
gradient only \cite{A5}. This model enables numerical computations of
the deformation of a smectic elastomer to be carried out without
modelling the microscopic length scale over which layer buckling
occurs. Resolving the length scale of the microstructure would require
the inclusion of spatial gradients in the deformation, for example
arising from Frank elastic energy. We will focus on using the coarse
grained free energy model for realistic geometries of tensile loading
of smectic elastomers that have been studied experimentally. A similar
programme has been successfully pursued for nematic elastomers, where
the free energy density is also non-convex. The resulting theoretical
predictions of microstructure
\cite{PhysRevE.66.061710,A3,desimonedolzmann2002} are in good
agreement with experiment \cite{finkelmannkundler1997,A7}.

\section{Model}

We will use the free energy density for a side-chain Sm-$A$ elastomer
derived in Ref. \cite{adams:05}. This has two contributions; the
energetic cost of changing the smectic layer spacing, and the entropic
term from stretching the underlying polymer network. It contains
parameters for the smectic layer modulus $B$, the rubber shear modulus
$\mu$, the polymer anisotropy $r$, and the reference state layer
normal $\vec{n}_0$. By using the high temperature isotropic state as
the reference state this free energy density can be simplified as shown
in appendix \ref{app:oldsmarelation}. It is also shown in appendix
\ref{app:oldsmarelation} that the simplified free energy density can
be approximated by the following expression
\begin{equation}
  W(\ten{F})= \textrm{Tr}\;\left(\ten{F}\cdot \ten{F}^T\right) + k(|\textrm{cof} \;
  \ten{F} \cdot \vec{n}_0| - q)^2
\label{eqn:wsmafe}
\end{equation}
where $\ten{F}= \nabla \vec{y}$ is the deformation gradient and
$\vec{y}$ is the displacement from the reference state, the cofactor
of $\ten{F}$ is denoted $\textrm{cof}\; \ten{F}= \ten{F}^{-T} $
(assuming $\textrm{det} \;\ten{F}=1$) and $W(\ten{F})$ has been made
dimensionless by dividing the original free energy density by $\half
\mu r^{1/3}$ \cite{A5}. The first term in Eq.~(\ref{eqn:wsmafe}) is
the entropic elasticity of the network, and the second is the smectic
layer compression term. We have disregarded an arbitrary additive
constant, and assumed that deformations are volume conserving,
i.e. $\textrm{det} \;\ten{F}=1$, in deriving this expression. The
constants $q$ and $k$ are given by
\begin{eqnarray}
  q &=& r^{-1/3} \left( 1+ \frac{\mu}{ B} (1-r)\right) \label{eqn:q}\\
  k &=& \frac{B}{\mu r^{2/3} q^3}. \label{eqn:k}
\end{eqnarray}
The free energy density of Eq.~(\ref{eqn:wsmafe}) is not convex, and
so the free energy of a homogeneous deformation can be lowered by the
formation microstructure, i.e.  a spatial variation in the deformation
gradient. Physically this microstructure corresponds to the buckling
of the smectic layers. The quasi-convex envelope of $W(\ten{F})$
provides a coarse grained free energy density that is optimised over
the possible microstructures, and is given by
\begin{eqnarray}
  W^{qc}(\ten{F})&=&\mathop{{\rm inf}}_{\vec{y} \in W^{1,\infty}_0}\left\{
    \frac{1}{|\Omega|}\int_\Omega
    W(\ten{F}+\nabla \vec{y}(x))dx \right.\nonumber\\
  &:&\left. \vec{y}(x) = 0 \;\;{\rm on } \;\;\partial \vphantom{\int}\Omega
  \right\},
\end{eqnarray}
where $\Omega$ denotes the volume of the domain, and
$\partial \vphantom{\int} \Omega$ denotes its boundary. An analytic
expression for the quasiconvex envelope of Eq.~(\ref{eqn:wsmafe}) was
derived in Ref. \cite{A5}, and we summarise it here. To write an
expression for $W^{qc}(\ten{F})$ we will need the largest singular
value of the matrix $\ten{F}$ denoted by
$\lambda_\textrm{max}(\ten{F})$, i.e.
\begin{equation}
  \lambda_\textrm{max}(\ten{F})= \textrm{sup} \left\{ | \ten{F} \cdot \vec{e}|: \vec{e} \in {\mathbb R}^3, |\vec{e}| = 1 \right\}.
\end{equation}
$W^{qc}(\ten{F})$ for the Sm-$A$ LCE model in Eq.~(\ref{eqn:wsmafe})
can be written in terms of the two convex functions of $\ten{F}$
\begin{eqnarray}
b&=&\lambda_\textrm{max}(\ten{F}\cdot \ten{P})^2\label{eqn:b}\\
d&=&| \textrm{cof}\; \ten{F} \cdot  \vec{n}_0|,\label{eqn:d}
\end{eqnarray}
where the matrix $\ten{P} = \Id - \vec{n}_0 \vec{n}_0^T$ projects out
the $\vec{n}_0$ component.
\begin{equation}
W^{qc}(\ten{F}) = \left\{\begin{array}[c]{cc}| \ten{F}\cdot \vec{n}_0|^2 + f(b,d)&\begin{array}{c}\textrm{det}\;\ten{F} = 1\\
 \textrm{ and } \\|\textrm{cof}\; \ten{F} \cdot \vec{n}_0|\leq q\end{array}
\\\\
\infty&\textrm{otherwise}\end{array}\right.
\label{eqn:wsmaqc}
\end{equation}
where
\begin{equation}
f(b,d) = \left\{ \begin{array}{cc}b + \frac{d^2}{b} + k (d-q)^2& d\geq \frac{k q b}{k b + 1}\\\\
b+\frac{k q^2}{q b + 1}&\begin{array}{c}b\geq q - \frac{1}{k}\\ \textrm{ and }\\ d\leq \frac{k q b}{k b +1}\end{array}\\\\
2 q - \frac{1}{k}& b \leq q - \frac{1}{k}
\end{array} \right..
\end{equation}
Eq.~(\ref{eqn:wsmaqc}) is a coarse grained model of a Sm-$A$ elastomer
that takes into account the formation of microstructure, without
resolving the fine-scale oscillations in the deformation gradient.

\subsection{Equilibrium}

To compare with experimental results it is convenient to work with
deformations relative to the low temperature equilibrium Sm-$A$
state. The system undergoes a volume conserving uniaxial deformation
as it is cooled from the isotropic state to the smectic state. This
uniaxial deformation along the layer normal minimises
Eq. (\ref{eqn:wsmaqc}). If we input the uniaxial deformation
\begin{equation}
\ten{F}_{0} = \left( \begin{array}{ccc}
1/\lambda_{0}^{2}& 0&0\\0&\lambda_{0} &0\\ 0&0&\lambda_{0}
\end{array}\right)
\end{equation}
into the total free energy $W^{qc}(\ten{F})$ of Eq.~(\ref{eqn:wsmaqc})
and then minimise it with respect to $\lambda_{0}$ we find the
equation
\begin{equation}
\frac{d}{d\lambda_{0}} \left[ k(\lambda_{0}^{2}-q)^{2} +2\lambda_{0}^{2} + \lambda_{0}^{-4} \right] =0.
\label{eqn:lambda_{0}}
\end{equation}
The value of $\lambda_{0}$ found by solving (\ref{eqn:lambda_{0}}) can
be used to convert deformations to start from the Sm-$A$ state as
follows
\begin{equation}
\ten{F} = \ten{F_{\text{Sm-$A$}}}\cdot\ten{F_{0}}
\label{eqn:equilibrium_transform}
\end{equation}
If we substitute this transformation into the free energy then the
uniaxial deformation $\ten{F}_0$ results in the scaling of $b$ and $d$
by $\lambda_{0}^{-2}$, and scaling of the term $|\ten{F}\cdot
\vec{n}_0|^2$ by $\lambda_{0}^{-4}$.
We will define the scaled quantities 
\begin{align}
\tilde{b} & =b/ \lambda_{0}^{2}  \label{eqn:b_tilde}\\
\tilde{d} & =  d/\lambda_{0}^{2} \label{eqn:d_tilde}
\end{align}
to describe $b$ and $d$ from the Sm-$A$ reference state.  The total
free energy with respect to the Sm-$A$ state (denoted with a tilde) is
\begin{eqnarray}
\nonumber
  \widetilde{W}^{qc}&&(\ten{F_{\text{Sm-$A$}}}) = \\
&&\left\{\begin{array}[c]{cc}
\begin{array}[c]{cc}\lambda_{0}^{-4}|\ten{F_{\text{Sm-$A$}}}\cdot \vec{n}_0|^2 \\+ f(\tilde{b},\tilde{d})\end{array}
&
\begin{array}{c}
\textrm{det}\;\ten{F_{\text{Sm-$A$}}}=1 \\
\textrm{and}\\
|\textrm{cof}\;\ten{F}_{\text{Sm-$A$}}\cdot \vec{n}_0|\leq q \lambda_0^2
\end{array}
\\\\
\infty&\textrm{otherwise}
\end{array}
\right.
\label{eqn:wscaled}
\end{eqnarray}
For the rest of this paper we will only refer to deformations with
respect to the Sm-$A$ reference state, so we will drop the subscript
on $\ten{F}_\textrm{Sm-$A$}$.

\begin{figure}[!htb]
\begin{center}
\includegraphics[width = 0.48\textwidth]{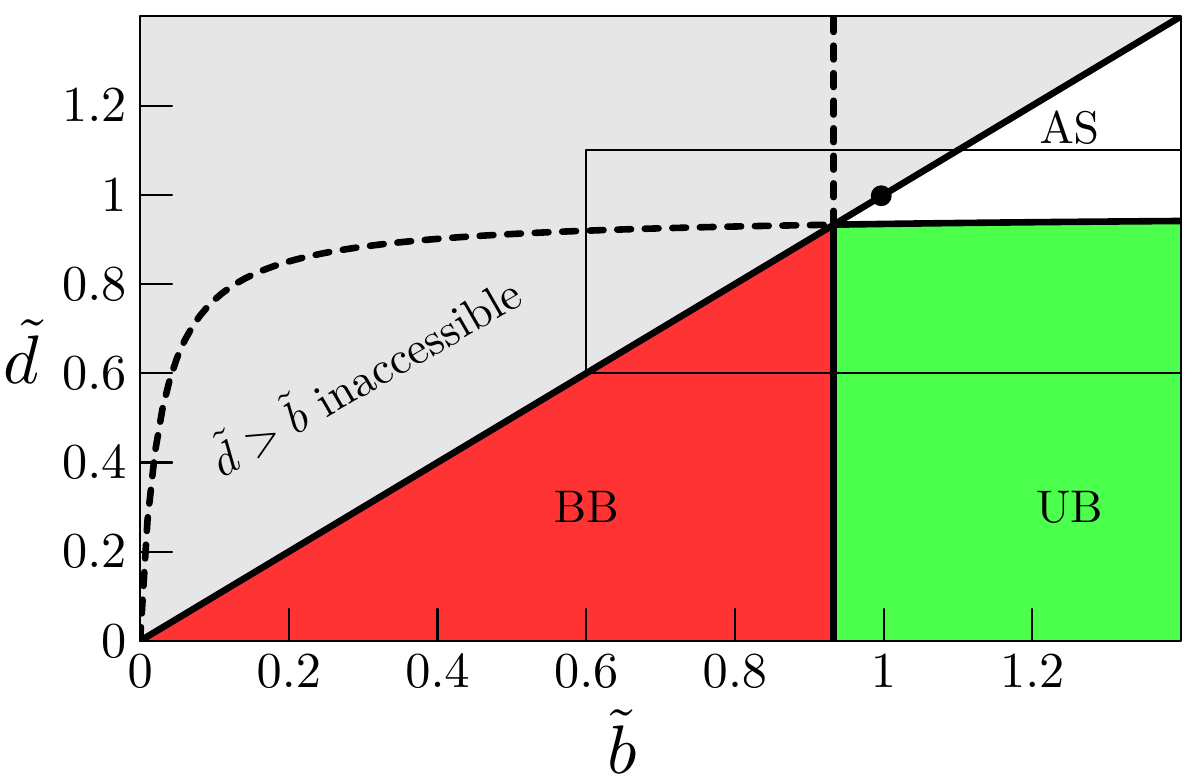}
\end{center}
\caption{The phase diagram of the Sm-$A$ LCE quasiconvex energy
  indicating the anisotropic solid (AS), uni-directional buckling (UB)
  and bi-directional buckling (BB) phases. The phase of the deformation
  is determined by $\tilde{b}$ and $\tilde{d}$ given in
  Eqs. (\ref{eqn:b_tilde}), (\ref{eqn:d_tilde}). The Sm-$A$
  equilibrium point is marked by a black circle.}
\label{fig:qcenergy}
\end{figure}

The phase diagram of the quasiconvex free energy is illustrated in
Fig.~\ref{fig:qcenergy}.
%
Note the region with $\tilde{d} > \tilde{b}$ is inaccessible for
volume conserving deformations. In the anisotropic solid (AS) phase
the quasiconvex free energy and the microscopic free energy are the
same. The energy is not lowered by the formation of microstructure,
and the smectic layers do not buckle. Hence the small angle X-ray
scattering pattern should show just one orientation of the layer
normal. In the uni-directional buckling (UB) phase the energy is
minimised by the formation of a simple laminate \cite{A5}. There are
two deformation gradients $\ten{F}_A$ and $\ten{F}_B$ that are rank
one connected, and whose suitably weighted average produces the
macroscopic deformation. The small angle X-ray scattering pattern will
contain two orientations of the layer normal corresponding to the
regions of $\ten{F}_A$ and $\ten{F}_B$. There should be no reduction
in X-ray scattering intensity if the beam is normal to the plane in
which the laminate forms. In the bi-directional buckling (BB) phase
there is no simple laminate that can achieve the optimal energy. A
higher order laminate must be formed \cite{A5}. Here the
microstructure contains an average of several different deformation
gradients. Physically buckling of the smectic layers in more than one
direction is possible, and it is expected that the small angle X-ray
scattering pattern will show a loss of intensity, indicating that some
smectic layers are rotated out of the scattering plane.

The three phases should be distinguishable using a crossed
polariser-analyser pair. The optical axis is parallel to the
director. The AS phase will appear dark when the polariser (or
analyser) is parallel to the optical axis, and have maximum brightness
when the polariser is at $45^\circ$ to the optical axis. In the BB
phase the director varies rapidly in both buckling directions, so it
will always be bright when viewed between the polariser-analyser. In
the UB phase the simple laminates associated with the uni-directional
layer buckling will be visible as striped domains, much like nematic
elastomers. We anticipate that both the BB phase and the UB phase will
be opaque, much like the striped domains in nematic elastomers (see
Refs. \cite{finkelmannkundler1997,A7}, and Fig. 8.10 of
\cite{warnerterentjev2007}) .

\subsection{Smectic layer buckling, finite extensibility and entanglements}

The Gaussian phantom chain network model neglects effects such as
finite extensibility of the polymer chains, and the entanglements of
chains with their neighbour. Several theoretical approaches have been
pursued to correct for these effects \cite{Treloar76,Deam76}.

The quasi-convex free energy in Eq.~(\ref{eqn:wsmaqc}) is formulated
on the assumption that an infinitely fine microstructure can be formed
at no energy cost. Energy terms involving gradients of the
deformation, arising through the Frank elastic cost of gradients in
the director will give rise to an interfacial energy
cost. Deformations perpendicular to the layer normal will distort the
buckled layers changing the interfacial energy.

$\widetilde{W}^{qc}$ is independent of $\tilde{b}$ and $\tilde{d}$ in
the BB phase, so it does not reproduce the Poisson's ratios of
$(\half, \half)$ seen in experiment. Motivated by the above
theoretical considerations, and to recover the experimentally observed
Poisson's ratio we will include an additional (convex) term that
physically relates to the non-Gaussian nature of the polymer chains, and
the deformation of the buckled layers. The magnitude of this
additional term arising from deforming the buckled layers can be
estimated through dimensional analysis as follows.

The free energy cost per unit area of interface in the microstructure
can be estimated as $\sqrt{K B}$ where $K$ is the Frank elastic
constant, and $B$ the liquid smectic modulus. The length scale of the
microstructure is given by the geometric mean of the sample size
parallel to the layer normal $L_x$ and the typical layer dimension,
i.e. $\sqrt{L_x\sqrt{\frac{K}{B}} }$. Using dimensional analysis we
can form an elastic modulus for the buckled layers by dividing these
two quantities
\begin{equation}
B \sqrt{\sqrt{\frac{K}{B}} \frac{1}{L_x}}.
\end{equation}
Note the buckled layer modulus goes to zero when $K=0$ as expected. A
more detailed calculation that produces a similar result for the
modulus is given in appendix \ref{app:modulusest}.

We will include in the energy a phenomenological Mooney-Rivlin type
term proportional to the second invariant of the Cauchy-Green strain
tensor $\ten{C} = \ten{F}^T \cdot \ten{F}$ \cite{Mooney40,Rivlin48}
\begin{eqnarray}
  \widetilde{W}_{MR}(\ten{F}) = \half  c_{MR} (\text{Tr}[\ten{C}]^{2}-\text{Tr}[\ten{C}\cdot\ten{C}])
\label{eqn:pert}
\end{eqnarray}
Note that the Mooney-Rivlin model is overly simplistic in assuming
that the derivatives of the energy with respect to the first, and
second invariants (denoted $A_1$ and $A_2$ respectively in
Eqs.~(\ref{eqn:A1}) and (\ref{eqn:A2})), $\frac{\partial W}{\partial
  A_{1}}$ and $\frac{\partial W}{\partial A_{2}}$, are constants, so
it does not realistically describe the uniaxial or biaxial stretching
of even isotropic rubbers
\cite{Treloar76,Gottlieb87,Kawamura01}. Consequently the values of
coefficients fitted to experiments are likely to be only approximate.

The total free energy 
\begin{equation}
  \widetilde{W}_\textrm{tot} = \widetilde{W}^{qc}+\widetilde{W}_{MR}
\label{eqn:tot}
\end{equation}
is altogether polyconvex \cite{neff2003}. The contribution to $c_{MR}$
from layer buckling in appropriate dimensionless units is
\begin{equation}
c_{MR} \sim \frac{2 B}{\mu r^{1/3}} \sqrt{\sqrt{\frac{K}{B}} \frac{1}{L_x}}.
\label{eqn:mrestimate}
\end{equation}

$\widetilde{W}_{MR}$ has a minimum when $\ten{F} = \Id$. If we substitute the
deformation
\begin{equation}
\ten{F} = \left( \begin{array}{ccc}
1& 0&0\\0&\lambda &0\\ 0&0&\frac{1}{\lambda}
\end{array}\right)
\end{equation}
into Eq.~(\ref{eqn:pert}) it produces the following
\begin{equation}
  \widetilde{W}_{MR}(\ten{F}) = 3 c_{MR} + 4 c_{MR} (\lambda-1)^2+ \mathcal{O} (\lambda-1)^3.
\end{equation}
Hence this additional term is minimal when the deformation in the two
transverse directions are equal. Consequently it will act to equalise
the Poisson's ratios, as seen in experiment.

Note that this additional term affects all the phases, not just the BB
phase. However it is not the dominant free energy term in the AS and
UB phases, so does not alter the physics of the model there.

\subsection{Model Parameters}

Our aim here is to model Sm-$A$ samples similar to those of Nishikawa
\cite{Nishikawa:97b,nishikawa1999}, hence we will use the material
parameters listed in Table \ref{tab:model_parameters} for the smectic
layer modulus $B$, the rubber shear modulus $\mu$, and a polymer
anisotropy $r$ appropriate for a prolate side chain
LCP. Eqs.~(\ref{eqn:q}),(\ref{eqn:k}) and (\ref{eqn:lambda_{0}}) can
then be used to find $q, k$ and $\lambda_0$.

We will use a value of the Mooney-Rivlin coefficient $c_{MR}=0.14$ in
finite element calculations. This can be estimated from
Eq.~(\ref{eqn:mrestimate}) with $L_x \sim 1$cm. Determination of this
value is discussed in \S \ref{sec:Finite Element Model}. However, it
is consistent with the work of Stannarius \textit{et al.}, who
performed mechanical experiments on Sm-$A$ LCE balloons and found
Mooney-Rivlin coefficients in the range $0< c_{MR}<0.1$
\cite{PhysRevE.65.041707}.

\begin{table}[h!tb]
\begin{tabular}{ l c }
  \hline
  Parameter (symbol) & Value \\ \hline
  $B$ & $6\times 10^6\text{Pa}$ \\
  $\mu$ & $10^{5}\text{Pa}$ \\
  $r$ & $2$ \\
  $K$ & $10^{-11}\text{N}$ \\\\
  $\lambda_{0}$ & $0.902$ \\
  $k$ & $48.43$ \\
  $q$ & $0.780$ \\\\
  $c_{MR}$ & $0.14$ \\
 \hline
\end{tabular}
\caption{Model Parameters}
\label{tab:model_parameters}
\end{table}

\section{Uniform Deformations}

To develop an intuition for the quasi-convex free energy in
Eq.~(\ref{eqn:wscaled}) we will now examine some uniform
deformations. Here we will assume that the layer normal is aligned
with the $\vec{x}$ direction, i.e. $\vec{n}_0 = (1,0,0)^T$.

\subsection{Elongation parallel to the layer normal}

An elongation parallel to the layer normal is described by
\begin{equation}
 \ten{F}_\parallel = 
\left( 
 \begin{matrix}
  \lambda & 0 & 0 \\
  0 & \frac{1}{\lambda^{\gamma}} & 0 \\
  0 & 0 & \frac{1}{\lambda^{1-\gamma}} 
 \end{matrix}
 \right),
\end{equation}
where the parameter $\gamma$ determines the Poisson's ratio of the
deformation. A value of $\gamma=\frac{1}{2}$ gives isotropic behaviour
in the directions perpendicular to the $\vec{n}_0$. A value of
$\gamma=1$ gives the anisotropic Poisson's ratios of $(1, 0)$.
\begin{figure}[!htb]
\begin{center}
\includegraphics[width = 0.48\textwidth]{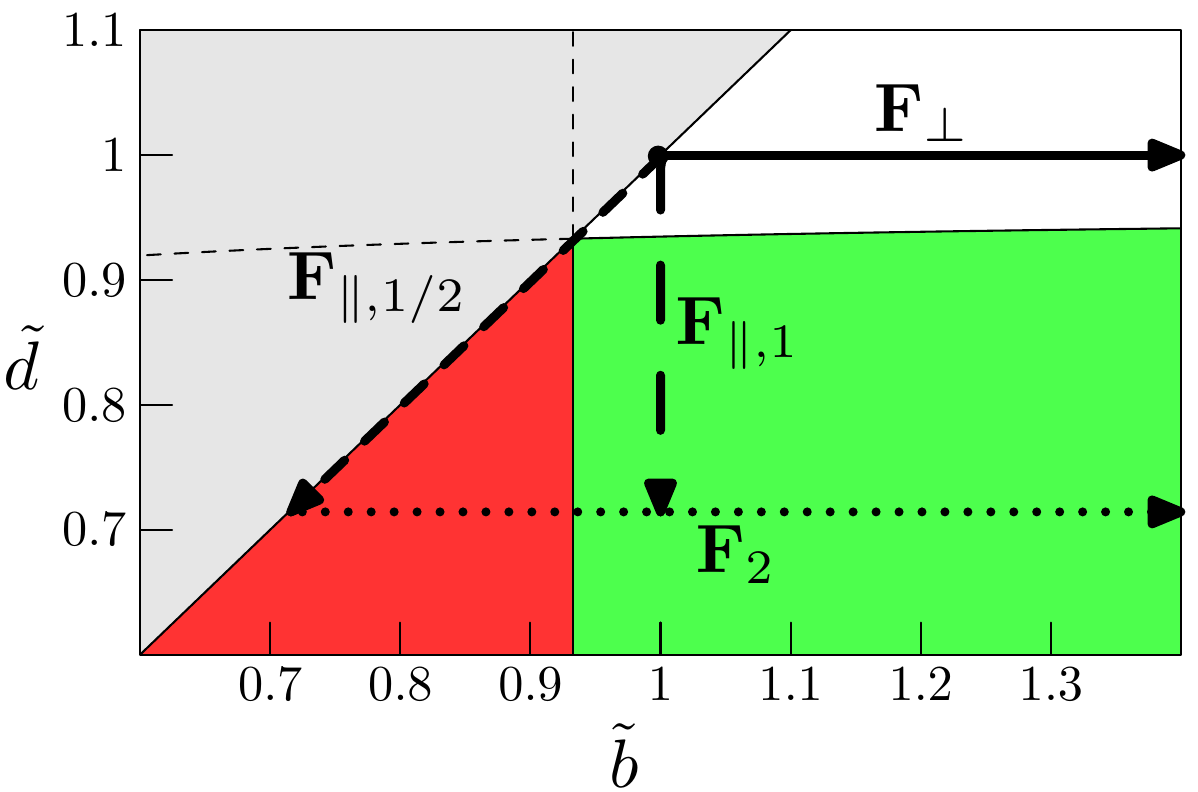}
\end{center}
\caption{Paths traversed in $\tilde{b}$ and $\tilde{d}$ on stretching
  parallel to $\vec{n}_0$ with (short dashed-line)
  $\gamma=\frac{1}{2}$ and (long dashed-line) $\gamma=1$.  The
  solid-line and the dotted-line are stretches perpendicular to
  $\vec{n}_0$, with the latter performed after an initial parallel to
  $\vec{n}_0$.}
\label{fig:b_d_diagram}
\end{figure}
Fig.~\ref{fig:b_d_diagram} shows the boxed area of the phase diagram
in Fig.~\ref{fig:qcenergy} and illustrates that when stretching
parallel to $\vec{n}_0$ with $\gamma=\half$ (labelled
$\ten{F}_{\parallel, 1/2}$ in Fig.~\ref{fig:b_d_diagram}) the
elastomer deformation follows the line $\tilde{b}=\tilde{d}$. The
system crosses from the AS to BB phase at a threshold deformation
$\lambda_\textrm{th}=\lambda_{0}^{2}(q-1/k)^{-1}$. By contrast, when
stretching parallel to $\vec{n}_0$ with $\gamma=1$ the elastomer
deformation follows the line of constant $\tilde{b}$ (labelled
$\ten{F}_{\parallel, 1}$ in the Fig.~\ref{fig:b_d_diagram}).

\begin{figure}[!htb]
\begin{center}
\includegraphics[width = 0.48\textwidth]{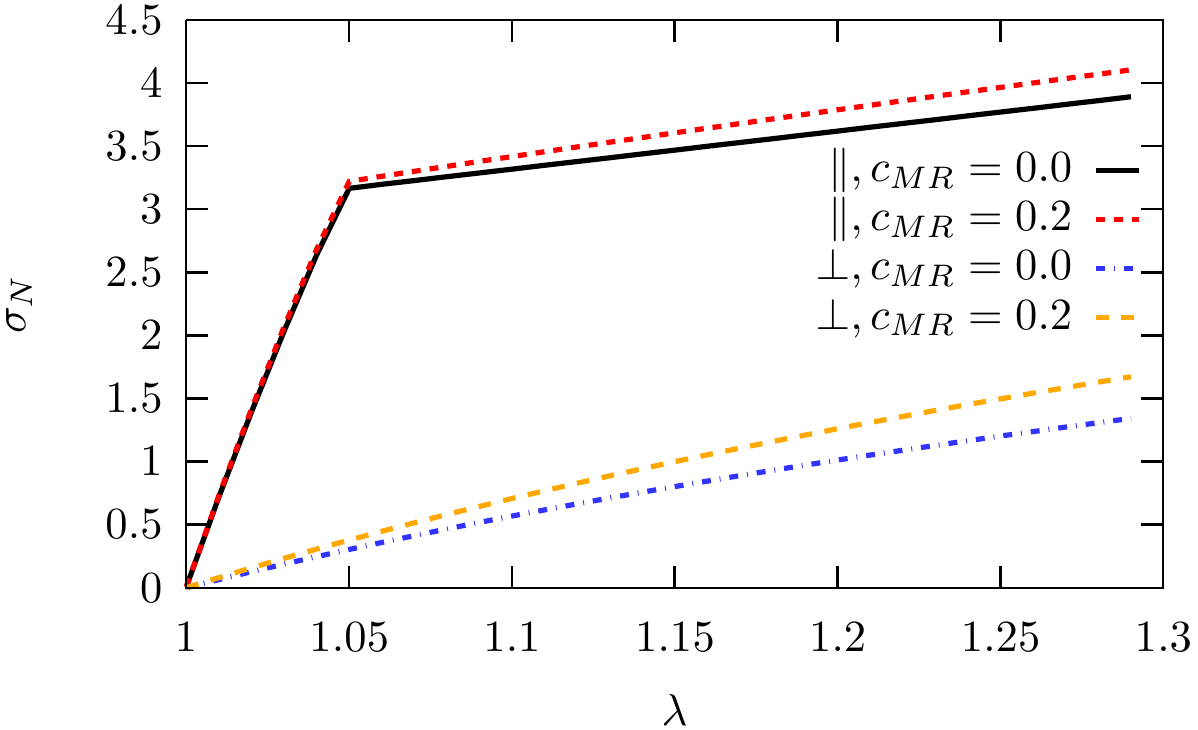}
\end{center}
\caption{The nominal stress $\sigma_N$ as a function of deformation
  $\lambda$ parallel to $\vec{n}_0$ with $\gamma=\half$, and
  perpendicular to $\vec{n}_0$.}
\label{fig:b_d_para_perp_stress}
\end{figure}

The nominal stress denoted $\sigma_N$, and measured in units of $\half
\mu r^{1/3}$ can be calculated by differentiating the scaled free
energy $\widetilde{W}_\textrm{tot}$ with respect to $\lambda$. The
nominal stress shows a dramatic reduction when the elastomer crosses
into the microstructured phases BB or UB. For example on the $\gamma =
\half$ trajectory the elastic modulus when the deformation begins is
\begin{equation}
k \frac{\lambda_0^4}{\lambda_\textrm{th}}+\frac{4}{\lambda_0^4} + 6 c_{MR}.
\end{equation}
This is dominated by the smectic layer modulus encoded in $k\gg
1$. After the threshold at $\lambda_\textrm{th}$ the modulus drops to
\begin{equation}
\frac{2}{\lambda_0^4} + \frac{6}{\lambda_\textrm{th}^4} c_{MR},
\end{equation}
i.e.  it is reduced by a factor of approximately $k$. This reduction
in the modulus is illustrated in Fig. \ref{fig:b_d_para_perp_stress}.

\subsection{Elongation  perpendicular to the layer normal}

An elongation perpendicular to the layer normal, $\vec{n}_0$, with
Poisson's ratios of $(1, 0)$, is described by
\begin{equation}
 \ten{F}_\perp = 
 \left( 
  \begin{matrix}
  1 & 0 & 0 \\
  0 & \lambda& 0 \\
  0 & 0 & \frac{1}{\lambda} 
 \end{matrix}
 \right).
\end{equation}
The trajectory of this deformation is along a line of constant
$\tilde{d}$, as shown in Fig. \ref{fig:b_d_diagram} (labelled
$\ten{F}_\perp$). The elastic modulus in this case is
\begin{equation}
8 \lambda_0^2 + 8 c_{MR}.
\end{equation}
The nominal stress $\sigma_N$ for this geometry is illustrated in
Fig.~\ref{fig:b_d_para_perp_stress}. There is no threshold in this
stress-strain curve, and no microstructure forms in this deformation
geometry.

\subsection{Two step deformation}

\begin{figure}[!htb]
\begin{center}
\includegraphics[width = 0.48\textwidth]{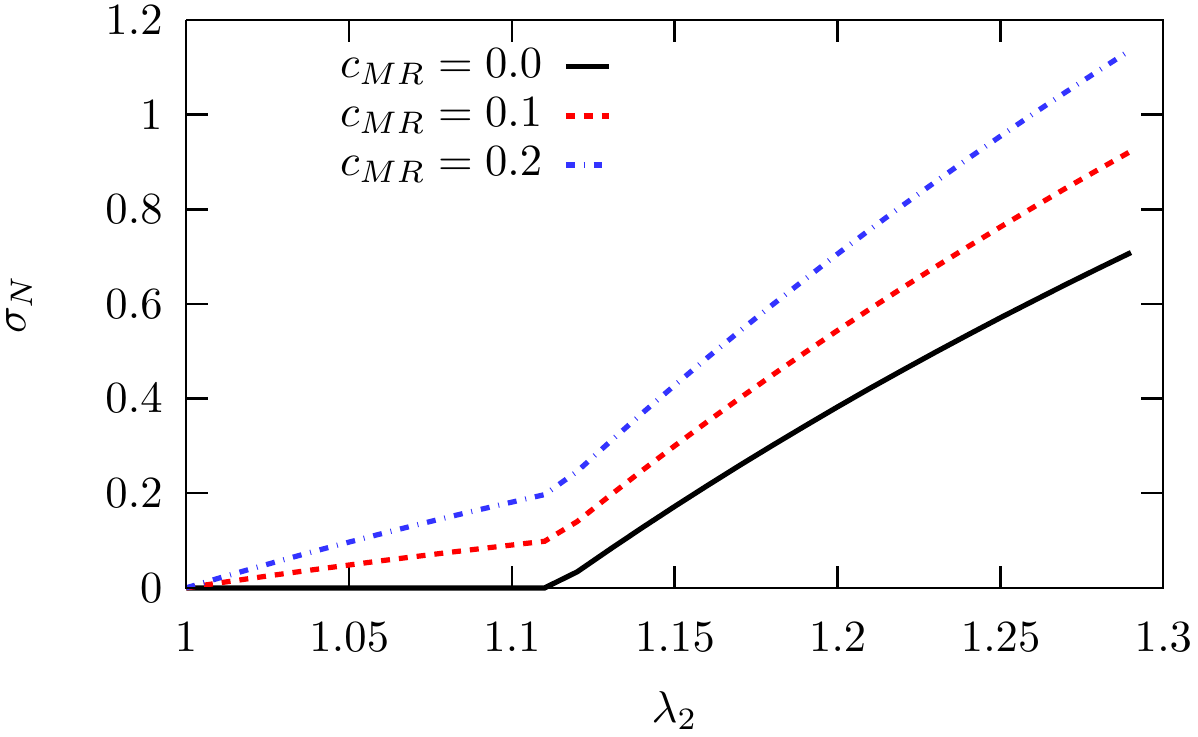}
\end{center}
\caption{Nominal stress as a function of deformation $\lambda_{2}$
  during the two stage deformation. The first stage is a deformation
  parallel to $\vec{n}_0$ of $\lambda_{1}=1.4$, followed by the
  perpendicular elongation $\lambda_{2}$.}
\label{fig:b_d_two_step_stress}
\end{figure}

A two stage deformation process first parallel to the layer normal by a
factor of $\lambda_1$, and then perpendicular to it by a factor
$\lambda_2$, defined in Eq.~(\ref{eqn:2stepF}), can be used to
experimentally determine the constant $c_{MR}$.
\begin{equation}
 \ten{F}_2 = 
 \left( 
  \begin{matrix}
  1 & 0 & 0 \\
  0 & \lambda_{2}& 0 \\
  0 & 0 & \frac{1}{\lambda_{2}} 
 \end{matrix}
 \right).
\left( 
 \begin{matrix}
  \lambda_{1} & 0 & 0 \\
  0 & \frac{1}{\sqrt{\lambda_{1}}} & 0 \\
  0 & 0 & \frac{1}{\sqrt{\lambda_{1}}} 
 \end{matrix}
 \right),
 \label{eqn:2stepF}
\end{equation}
The trajectory of this deformation is illustrated in
Fig.~\ref{fig:b_d_diagram}. The first stage follows
$\ten{F}_{\parallel,1/2}$, and the second stage is labelled
$\ten{F}_2$.  The first stage of deformation proceeds the system moves
along the line $\tilde{b}=\tilde{d}$, thus crossing from the AS to BB
phase. During the second deformation stage the system moves along a
line of constant $\tilde{d}$, crossing from the BB to UB phase. The
nominal stress during the second stage is shown in
Fig. \ref{fig:b_d_two_step_stress}. If $c_{MR}$ is zero then the
deformation is perfectly soft within the BB phase. This is an
intrinsic property of $\widetilde{W}^{qc}(\ten{F})$ which is altered
by the addition of $\widetilde{W}_{MR}$. Physically this reflects the
fact that there is an energetic cost to deforming buckled layers,
which rules out perfectly soft deformation.
At the start of the $\lambda_2$ deformation the elastic modulus is
given by 
\begin{equation}
8 c_{MR}\lambda_1
\end{equation}
i.e. it is entirely due to the additional Mooney-Rivlin term, so can
be used to experimentally measure this additional constant. Once the
trajectory of the deformation enters the UB phase the stiffness
increases to
\begin{equation}
  8 q + 8 /(q k ^2) - 16 /k +8 c_{MR}\lambda_1.
\end{equation}

\subsection{Elongation at an angle to the layer normal}

\begin{figure}[!htb]
\begin{center}
\includegraphics[width = 0.48\textwidth]{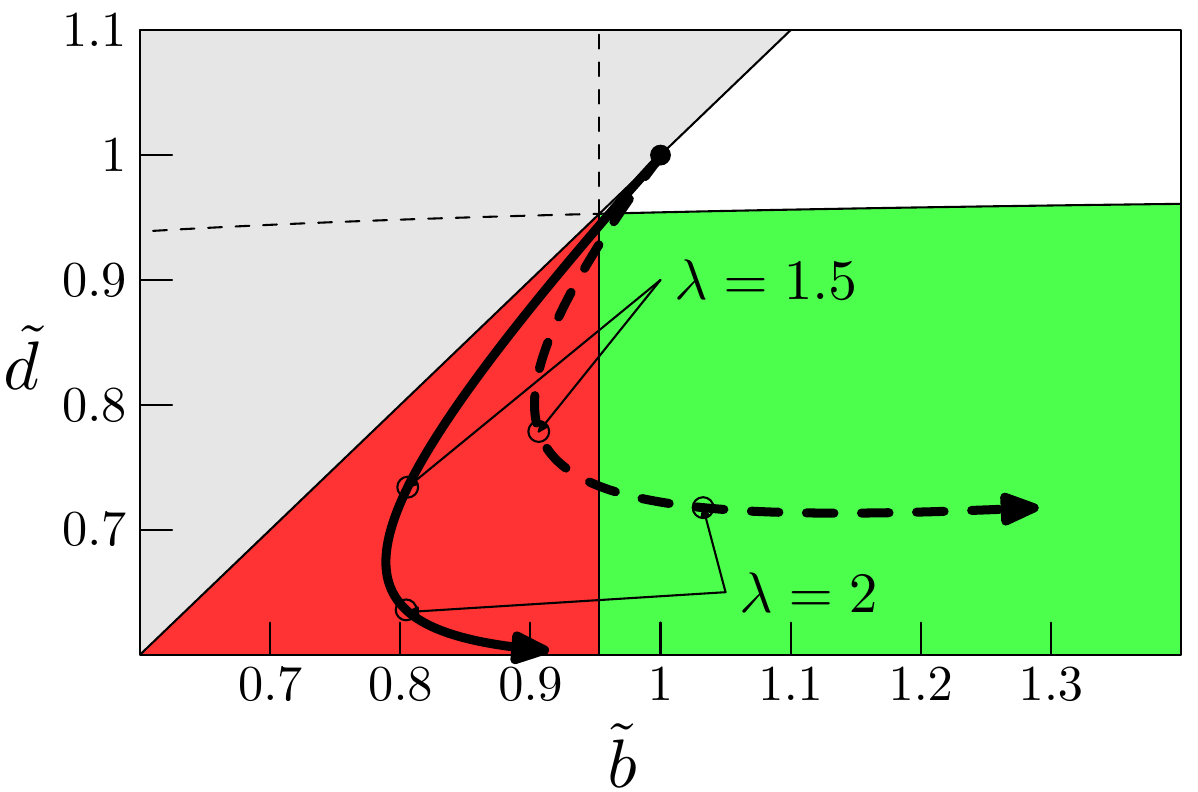}
\end{center}
\caption{The trajectories on the phase diagram for elongations at an
  angle of $\theta = 17^\circ$ (solid line) and $\theta =
  23^\circ$ (dashed line) to $\vec{n}_0$.  The maximum
  deformation shown in each case corresponds to $\lambda = 2.5$.}
\label{fig:b_d_angle}
\end{figure}

Elongation of the elastomer at an angle $\theta$ to the layer normal
can be represented by the deformation 
\begin{equation}
\ten{F} = 
\left( \begin{matrix}
  \lambda & 0 & 0 \\
  0 & \frac{1}{\sqrt{\lambda}} & 0 \\
  0 & 0 & \frac{1}{\sqrt{\lambda}} 
 \end{matrix} \right)
\cdot 
\left(
 \begin{matrix}
\cos \theta & \sin \theta & 0\\
- \sin \theta & \cos \theta & 0\\
0&0&1
 \end{matrix}
 \right).
\end{equation}
Two trajectories for this type of deformation are shown on the phase
diagram in Fig.~\ref{fig:b_d_angle} for $\theta = 17^\circ$ and
$23^\circ$.
Elongation at an angle to the layer normal results in a rapid rotation
of the layer normal away from the stretch axis. The lowest free energy
of the system for larger rotation angles is in the UB phase, as
illustrated by the trajectory of the deformation.

\section{Finite Element Model}
\label{sec:Finite Element Model}
\subsection{Method}

The free energy in Eq.~(\ref{eqn:tot}) derived from
Eqs.~(\ref{eqn:wscaled}) and (\ref{eqn:pert}) has one direction of
anisotropy $\vec{n}_{0}$. It can be written using the following
invariants of the Cauchy-Green strain tensor
$\ten{C}=\ten{F}^{T}\cdot\ten{F}$.
\begin{align}
A_{1} & = \text{Tr}[\ten{C}]\label{eqn:A1}\\
A_{2} & = \half(\text{Tr}[\ten{C}]^{2}-\text{Tr}[\ten{C}\cdot\ten{C}])\label{eqn:A2}\\
A_{3} & = \text{det}[\ten{C}]\\
A_{4} &= \vec{n}_0\cdot\ten{C}\cdot\vec{n}_0\\
A_{5} &= \vec{n}_0\cdot\ten{C}\cdot\ten{C}\cdot\vec{n}_0.
\end{align}
The parameters $\tilde{b}$ and $\tilde{d}$ can be rewritten as
\begin{align}
\tilde{b} & = \frac{A_{1} - A_{4} + \sqrt{(A_{1} + A_{4})^2 - 4 (A_{2} + A_{5})}}{2} \label{eqn:b_invariants}\\ 
\tilde{d} & = \sqrt{A_{2} + A_{5} - A_{1} A_{4}}.
\end{align}
The Mooney-Rivlin term can be rewritten as
\begin{equation}
W_{MR}(\ten{F}) = c_{MR} A_{2}.
\end{equation}

Some care must be taken in treating these expressions numerically.
Firstly in Eq.~(\ref{eqn:b_invariants}) the two terms $(A_{1} +
A_{4})^{2}$ and $4(A_{2} + A_{5})$ are typically close together. This
subtractive cancellation can lead to large numerical errors. Secondly
we require the derivatives of the free energy to compute the stresses
in the material. Differentiating the square root expression in
Eq.~(\ref{eqn:b_invariants}) gives an expression that diverges when
$(A_{1} + A_{4})^{2}=4(A_{2} + A_{5})$. It is useful to smooth the
divergence in this expression by adding a small value $\epsilon \sim
10^{-5}$ to the contents of the square root.

The material energy, $\widetilde{W}_\textrm{tot}(\ten{F})$, was
implemented in the commercial finite element package \texttt{Abaqus 6.10}
\cite{abaqus6.10} by writing a \texttt{UANISOHYPER\_INV} subroutine
for the standard implicit integration scheme.  The numerical method in
this routine is based on previous work implementing invariant based
elasticity \cite{Weiss96,Kaliske2000}.  Incompressibility is enforced
within this code by specifying \texttt{type=incompressible} in the
material definition. The anisotropy parameter \texttt{local
  directions=1} is specified, with the local direction defined as
$\vec{n}_0$.

Rigid clamping boundary conditions were used on the end faces of the
elastomer. In \texttt{Abaqus} these constraints are implemented as
pinned displacement boundary conditions, e.g. \texttt{U1=0.64,U2=0}
and \texttt{U3=0} at the mobile clamp. Experimentally an alternative
to rigid clamping is to secure the ends of the elastomer with tape,
which allows a contraction in thickness of the elastomer at the clamp.
Simulations using tape-like boundary conditions produce very similar
stress-strain curves to rigid clamping with a slight difference in
microstructure near the clamps.

The elastomer was deformed by moving one of the clamps to achieve a
total deformation of $\lambda = 1.4$. The step size increment was
fixed at $5 \times 10^{-3}$.

\subsection{Mesh Verification}

Initial tests of the \texttt{UANISOHYPER\_INV} subroutine were
conducted on a single C3D8H (8-node linear brick hybrid) element.
These showed that the model is correctly equilibrated, as no stresses
are present at zero deformation.  When stretching parallel to
$\vec{n}_0$ the expected stress-strain curve was
reproduced. Integration points undergo a transition from the AS to BB
phase at the correct threshold strain.

The subroutine was then tested with C3D8RH (reduced-integration) and
C3D20H (twenty-node) elements and it was confirmed that the results
were independent of the element-type.

The thin film was represented using uniform meshes with between $800$
($40 \times 20 \times 1$) and $32,000$ ($200 \times 160 \times 1$)
elements. These meshes were observed to achieve equivalent
results. Computations were also performed using biased meshes, which
achieved stress solutions within $0.5\%$ of uniform meshes. Equivalent
results were also obtained with thicker meshes ($100 \times 50 \times
5$).

The results presented in the following sections were obtained using a
rectangular uniform mesh of $5000$ ($100\times 50 \times 1$) C3D8H
elements.

\subsection{Parameters}

Motivated by the work of Nishikawa \textit{et al.}
\cite{nishikawa1999} we will start by investigating a sample
consisting of a rectangular cuboid of dimensions $1.6\text{cm} \times
1.0\text{cm} \times 500 \mu \text{m}$.

\begin{figure}[!h]
\begin{center}
  \includegraphics[width =
  0.48\textwidth]{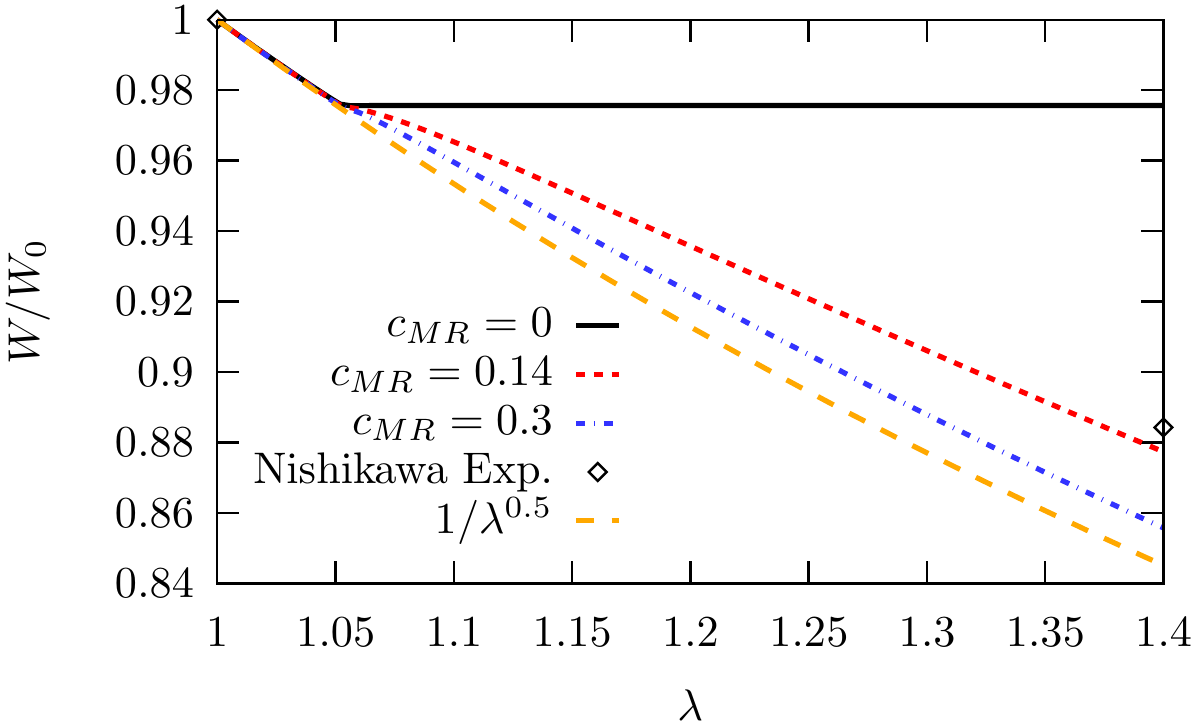}
\end{center}
\caption{Deformation across the width of the sample in the target
  state as a function of the deformation applied parallel to
  $\vec{n}_0$.}
\label{fig:alpha}
\end{figure}

The value of the layer buckling term, $c_{MR}$, can be estimated by
examining its effect on the fractional change of the width of the
sample, $W/W_0$ when stretching parallel to $\vec{n}_0$. The width of
the middle of the sample was measured as a function of
deformation. Fig.~\ref{fig:alpha} shows that if $c_{MR}=0$ the width
of the sample remains constant above the threshold. A value of
$c_{MR}=0.14$ successfully approximates the deformed state seen in
Fig.~4. of Ref. \cite{nishikawa1999}.
The other parameters used in the finite element calculations are as
presented in Table~\ref{tab:model_parameters}.

\section{Results and Discussion}

\subsection{Elongation parallel and perpendicular to the layer normal}

\begin{figure}[h!tb]
\begin{center}
  \includegraphics[width = 0.48\textwidth]{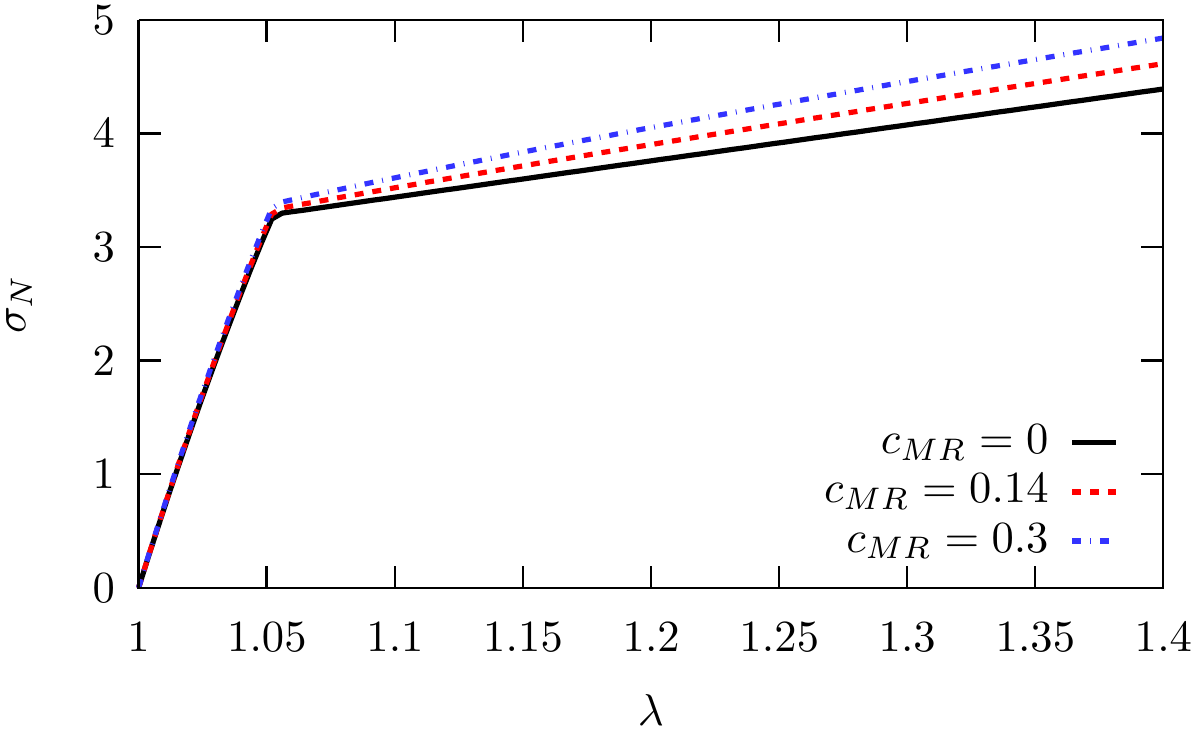}
\end{center}
\caption{The nominal stress as a function of deformation parallel to
  $\vec{n}_0$ for different values of $c_{MR}$.}
 \label{fig:stress_strain_parallel_perturbation}
\end{figure}

\begin{figure}[!htb]
\begin{center}
 \includegraphics[width = 0.48\textwidth]{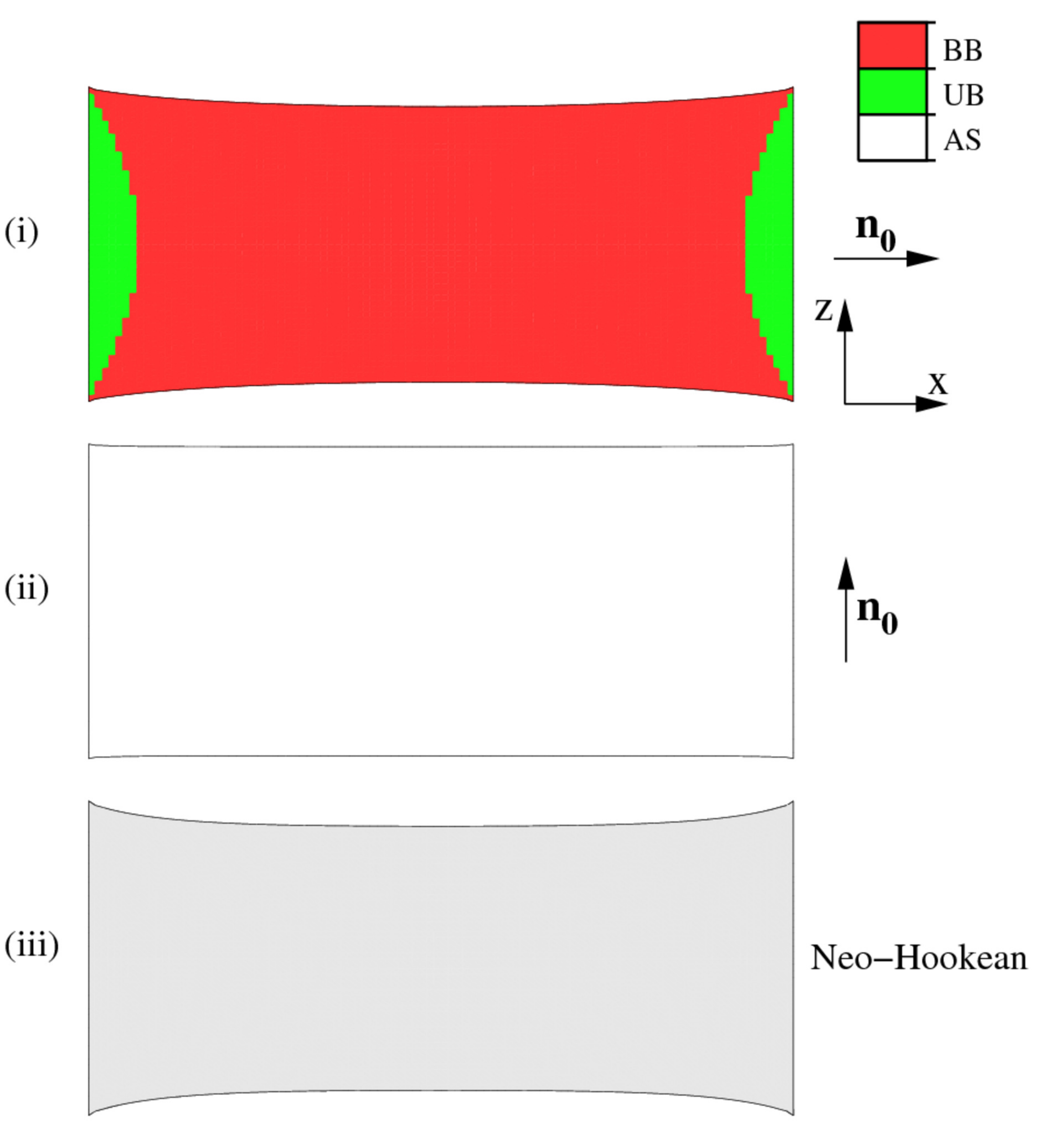}
\end{center}
\caption{Microstructure distribution when deforming (i) parallel to
  $\vec{n}_0$ and (ii) perpendicular to $\vec{n}_0$, shown at a
  deformation of $1.4$. (iii) An isotropic Neo-Hookean sample, with
  free energy $W(\ten{F}) = C_{1}(A_{1}-3)+
  \frac{1}{D_{1}}(A_{3}-1)^{2}$, where $C_{1}=2$ and
  $D_{1}=10^{-6}$.}
\label{fig:microstructure_neohookean}
\end{figure}

The stress-strain curve for deformation parallel to $\vec{n}_0$ is
shown in Fig.~\ref{fig:stress_strain_parallel_perturbation}. This
curve, obtained from finite element modelling, is in agreement with the
stress-strain curve obtained for a uniform deformation shown in
Fig.~\ref{fig:b_d_para_perp_stress}. 
The spatial distribution of phase of the sample is shown in
Fig.~\ref{fig:microstructure_neohookean}(i).  The bulk of the sample
is in the BB phase, however the UB phase is present in the vicinity of
the clamps. Near the clamps the elastomer is constrained in a way that
prevents isotropic deformation, meaning they tend to form UB
microstructure rather than BB microstructure. The shape of the
deformed sample is similar to that of the isotropic Neo-Hookean
elastomer shown in Fig.~\ref{fig:microstructure_neohookean}(iii).

On deforming the sample perpendicular to $\vec{n}_0$ no buckled
microstructure forms, as shown
Fig.~\ref{fig:microstructure_neohookean}(ii).  This behaviour is
consistent with the uniform deformation case shown in
Fig.~\ref{fig:b_d_diagram}. The layer spacing is constant and the
sample deforms with Poisson's ratios of $(1,0)$.

\subsection{Elongation at an arbitrary angle to the layer normal}

\begin{figure}[h!tb]
\begin{center}
\includegraphics[width = 0.43\textwidth]{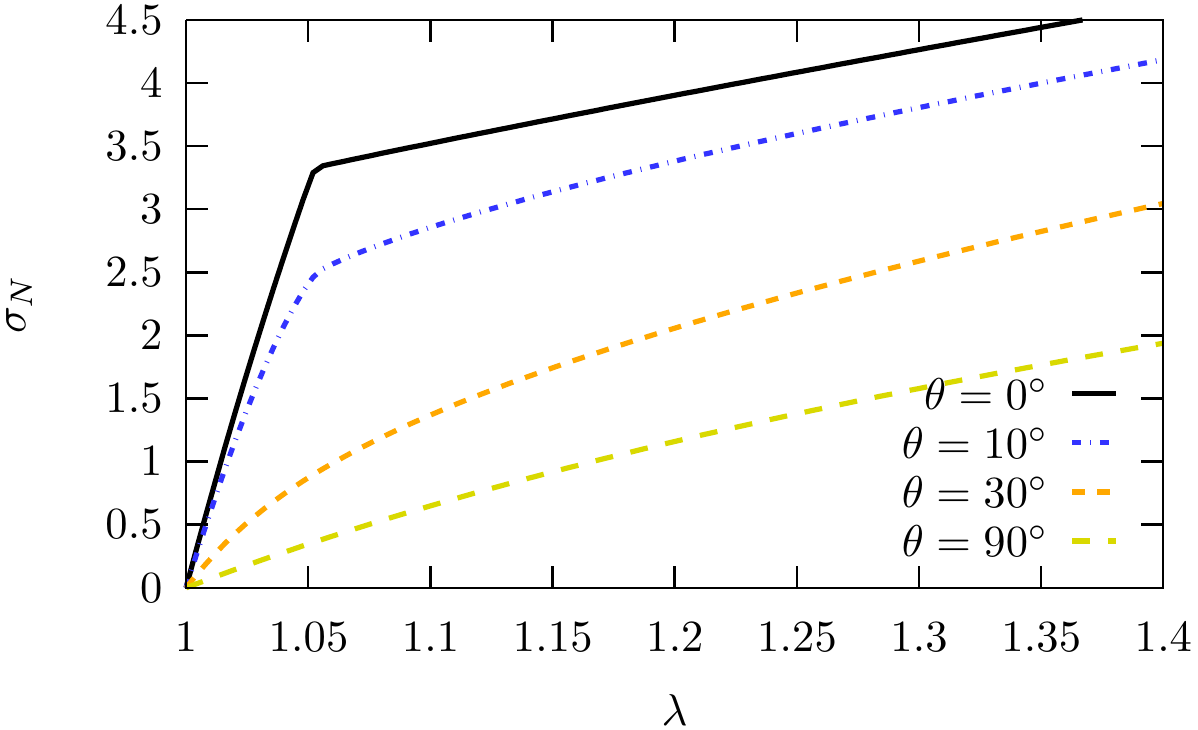}
\end{center}
\caption{The nominal stress as a function of deformation, where
  $\vec{n_{0}}$ is oriented in the plane of the film at an angle
  $\theta$ to the elongation axis.}
\label{fig:nominal_stress_strain_theta}
\end{figure}

\begin{figure}[h!tb]
\begin{center}
  \includegraphics[width =
  0.4\textwidth]{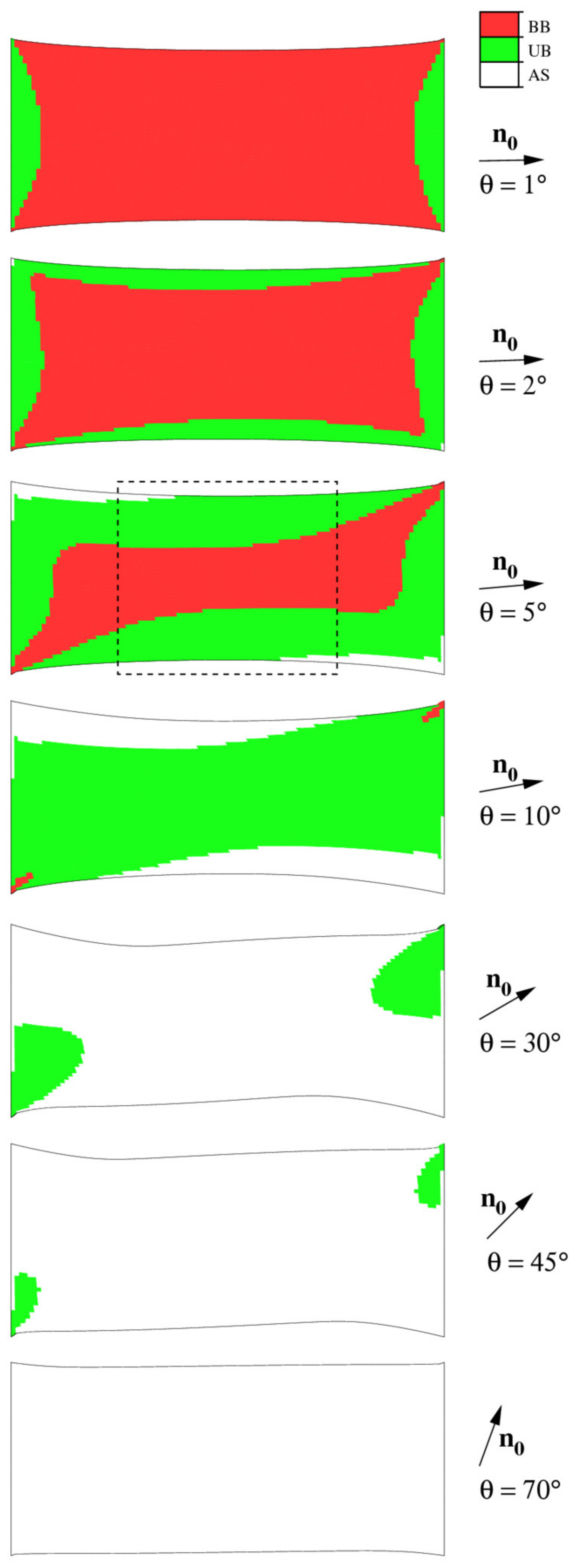}
\end{center}
\caption{Microstructure distribution for elongation at $1^{\circ}$,
  $2^{\circ}$, $5^{\circ}$, $10^{\circ}$, $45^{\circ}$ and
  $70^{\circ}$ to $\vec{n_{0}}$, shown at a deformation of
  $\lambda=1.4$. The dashed region is explored in more detailed in
  Fig. \ref{fig:sheared_microstructure}.}
\label{fig:microstructure_angles}
\end{figure}

The stress-strain behaviour for elongations at various angles to
$\vec{n_{0}}$ are shown in Fig. \ref{fig:nominal_stress_strain_theta}
for an elastomer with the same aspect ratio as those of Nishikawa
\textit{et al.}
For elongations within $\sim 10^{\circ}$ of $\vec{n_{0}}$ the
stress-strain curve still resembles that of the parallel case.
However for elongations at $\sim 30^{\circ}$ and above there is no
longer a well defined threshold transition to a lower modulus. The
corresponding spatial distribution of microstructure for elongations
at various angles to $\vec{n_{0}}$ are shown in
Fig.~\ref{fig:microstructure_angles}. These results show that
elongations at an angle within $\sim 1^{\circ}$ of $\vec{n_{0}}$
result in the BB phase forming in the bulk of the sample, with UB
phase at the clamps. Note that for angles above $\sim 30^{\circ}$
there is no percolation of the strip of the UB phase across the
sample. This coincides with the disappearance of the threshold in the
stress-strain response.
At a stretching angle of $2^{\circ}$ the UB phase forms at the free
edges of the sample.  The formation of UB microstructure is
accompanied by $\lambda_{xz}$ shears present in these regions of the
sample.

\begin{figure}[h!tb]
\begin{center}
  \includegraphics[width =
  0.48\textwidth]{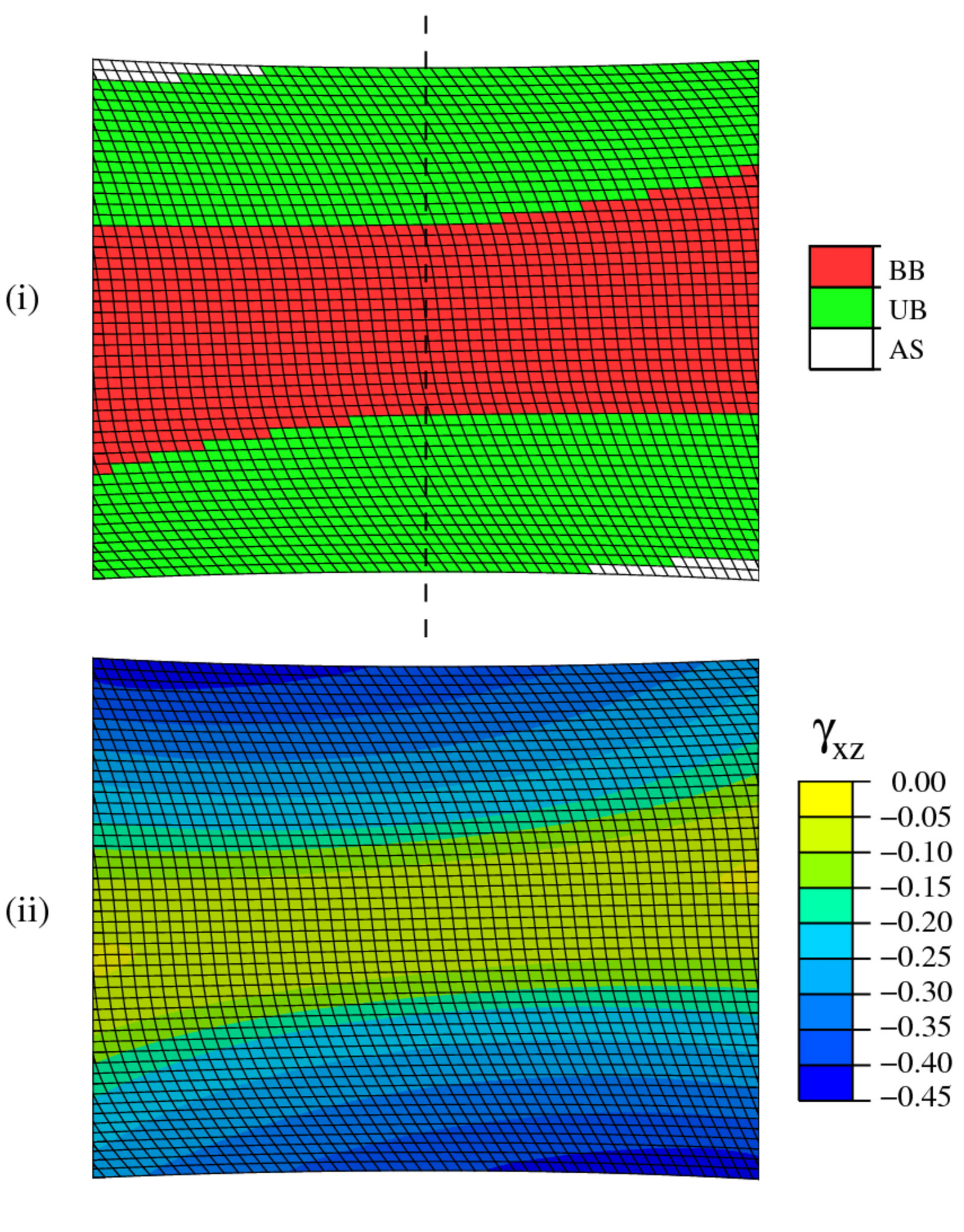}
\end{center}
\caption{(i) Spatial microstructure distribution and (ii)
  $\gamma_{xz}$ engineering shear strain, for the dashed region of
  Fig. \ref{fig:microstructure_angles}.  }
\label{fig:sheared_microstructure}
\end{figure}

We will now examine the deformation of the sample with a $5^\circ$
inclination of the layer normal in more depth. The phase distribution
and the shear deformation are shown in
Figs.~\ref{fig:sheared_microstructure} (i) and (ii) respectively. The
deformation of the mesh shows the shear deformation of the
elements. Only the weakly sheared, central area of the sample is in
the BB phase.  Strong shears result in a transition from BB to UB
phase. The transition occurs at $\lambda_{xz}\sim 0.5$ for an imposed
deformation of $\lambda_{xx}=1.4$, or equivalently an engineering
shear strain of $\gamma_{xz}=(\lambda_{xz}+\lambda_{zx})/2\sim 0.25$.

\begin{figure}[!htb]
\begin{center}
\includegraphics[width = 0.48\textwidth]{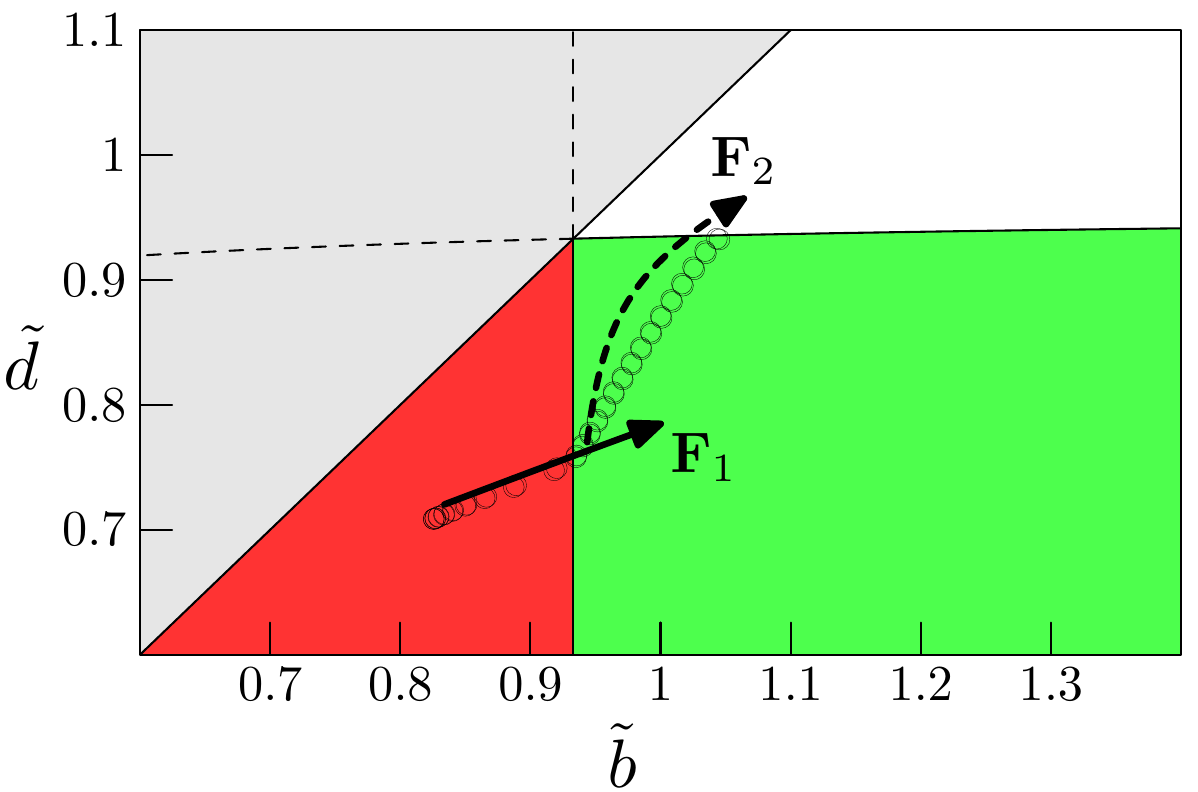}
\end{center}
\caption{The phase of the elements across the sample in
  Fig. \ref{fig:sheared_microstructure} (i) in the $(\tilde{b},
  \tilde{d})$ phase space (open circles). The trajectory $\ten{F}_1$
  is a deformation with $\gamma = 0.73$, $\theta = 5^\circ$,
  $\lambda_1 = 1.4$, and $\lambda_{xz} = 0$ to $0.4$ (solid line). The
  trajectory $\ten{F}_2$ is a deformation with $\gamma = 0.1$, $\theta
  = 5^\circ$, $\lambda_1 =1.4$ and $\lambda_{xz} = 0.4$ to $0.7$
  (dashed line).  }
\label{fig:b_d_shear}
\end{figure}

We can understand these results, and the transformation of the sample
phase by considering a deformation at an angle $\theta$ to $\vec{n}_0$
consisting of an elongation $\lambda_1$, and a shear $\lambda_{xz}$,
\begin{equation}
 \ten{F} = 
\left( 
 \begin{matrix}
  \lambda_{1} & 0 & \lambda_{xz} \\
  0 & \frac{1}{\lambda_{1}^\gamma} & 0 \\
  0 & 0 & \frac{1}{\lambda_{1}^{\gamma -1}} 
 \end{matrix}
 \right)\cdot 
\left(
 \begin{matrix}
\cos \theta & \sin \theta & 0\\
- \sin \theta & \cos \theta & 0\\
0&0&1
 \end{matrix}
\right).
\end{equation}
The state of the elements from the slice across the sample in
Fig. \ref{fig:sheared_microstructure} (i) in the $(\tilde{b},
\tilde{d})$ phase space is shown in Fig.~\ref{fig:b_d_shear}.
As we cross the centre of the sample, the elements are in the BB
phase. The elements are subjected to an increased amount of shear,
which is illustrated by the trajectory labelled $\ten{F}_1$ in
Fig.~\ref{fig:b_d_shear}. Once the edge of the BB phase is reached the
thickness of the sample increases, and the sample transitions to the
UB phase. As the UB phase consists of buckling in only one direction,
it is thicker in the direction perpendicular to the plane in which the
microstructure laminates are formed. This is illustrated by the
trajectory $\ten{F}_2$ in Fig.~\ref{fig:b_d_shear}.

\subsection{Aspect Ratio}

\begin{figure}[h!tb]
\begin{center}
  \includegraphics[width =
  0.48\textwidth]{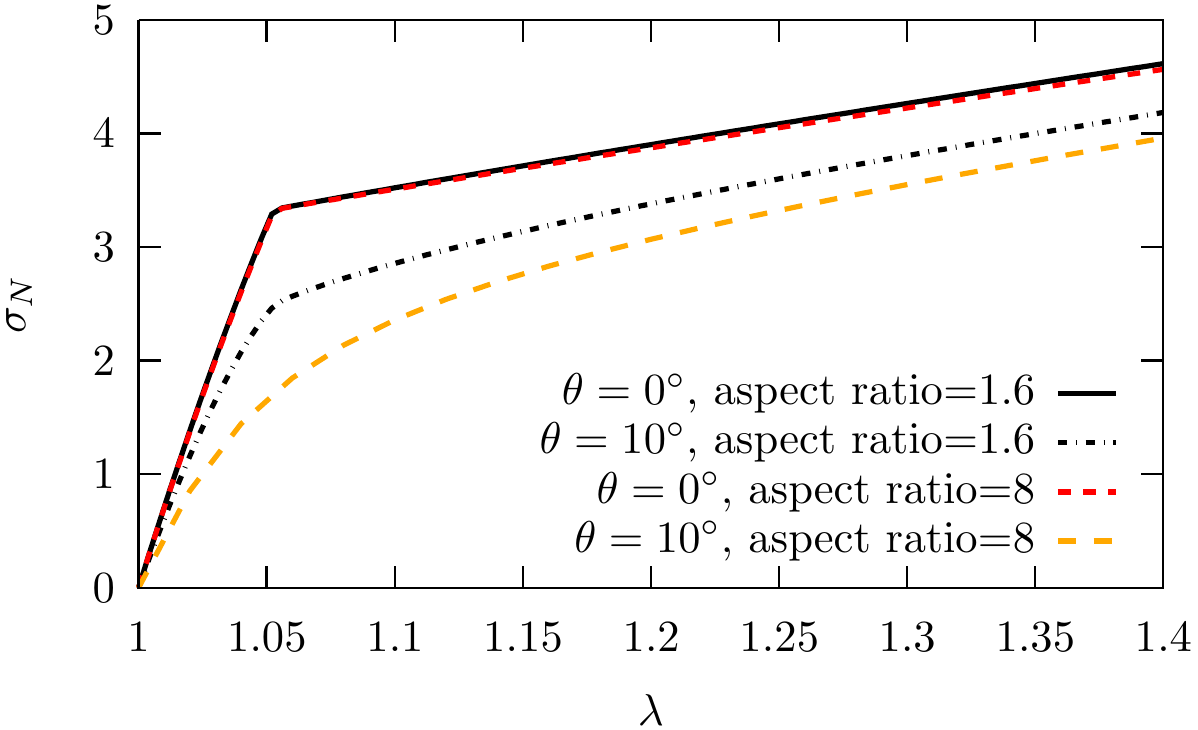}
\end{center}
\caption{Nominal stress as a function of deformation for elongation at
  $0^{\circ}$ and 10$^{\circ}$ to $\vec{n_{0}}$, for aspect ratios of
  $1.6$ and $8$.}
\label{fig:nominal_stress_strain_theta_aspect_ratio}
\end{figure}

So far we have only considered samples with the same aspect ratio as
Nishikawa \textit{et al.} \cite{nishikawa1999}. Other work on Sm-$A$
elastomers has used very different sample aspect ratios such as Komp
\textit{et al.} \cite{MARC:MARC200600640}. The finite element results
shown in Fig.~\ref{fig:nominal_stress_strain_theta_aspect_ratio} show
that varying the length-to-width ratio of the sample at constant film
thickness alters the stress-strain curves obtained when at a small
angle to $\vec{n}_0$, but produce the same stress strain curves when
stretching exactly parallel to $\vec{n}_0$.

\begin{figure}[h!tb]
\begin{center}
\includegraphics[width = 0.48\textwidth]{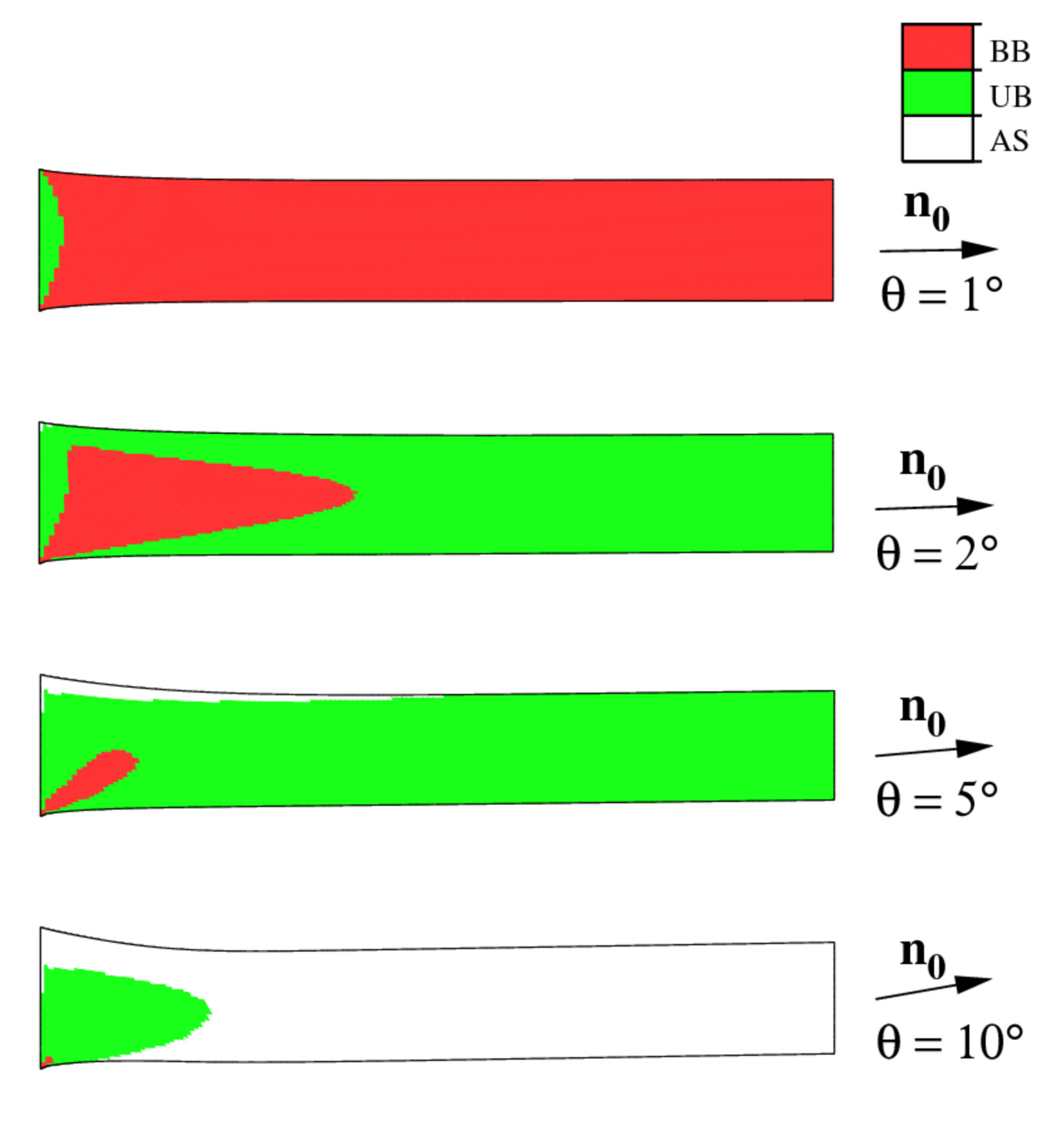}
\end{center}
\caption{Microstructure distribution when stretching at 1$^{\circ}$,
  2$^{\circ}$, 5$^{\circ}$ and 10$^{\circ}$ to $\vec{n_{0}}$ at a
  strain of 0.4.  The sample dimensions are $8.0 \text{cm} \times
  1.0\text{cm} \times 500 \mu \text{m}$, which is an aspect ratio of
  8. Only half of the samples are depicted here.}
\label{fig:8-1-0.05_microstructure_arbitrary}
\end{figure}

The spatial microstructure distribution is highly sensitive to the
aspect ratio. Fig. \ref{fig:8-1-0.05_microstructure_arbitrary} shows
the microstructure distribution in a sample with an aspect ratio of
$8$. When compared to Fig. \ref{fig:microstructure_angles}, where the
aspect ratio is $1.6$ it can be seen that the larger aspect ratio
reverts to the AS phase for smaller angles of inclination of the
deformation to the layer normal.
Qualitatively this is because a smaller fraction of the sample is
taken up by the end region near the clamps as the aspect ratio
increases. Hence the layer normal is less constrained in its rotation
by these end regions, and can adopt the lowest energy orientation
rotated away from the elongation axis. For the aspect ratio of $8$ an
inclination of as little as $2^\circ$ results in the sample forming
the UB phase rather than the BB phase.  This may make it difficult to
experimentally observe BB microstructure in high aspect ratio samples
by stretching parallel to $\vec{n_{0}}$.

\begin{figure}[h!tb]
\begin{center}
\includegraphics[width = 0.48\textwidth]{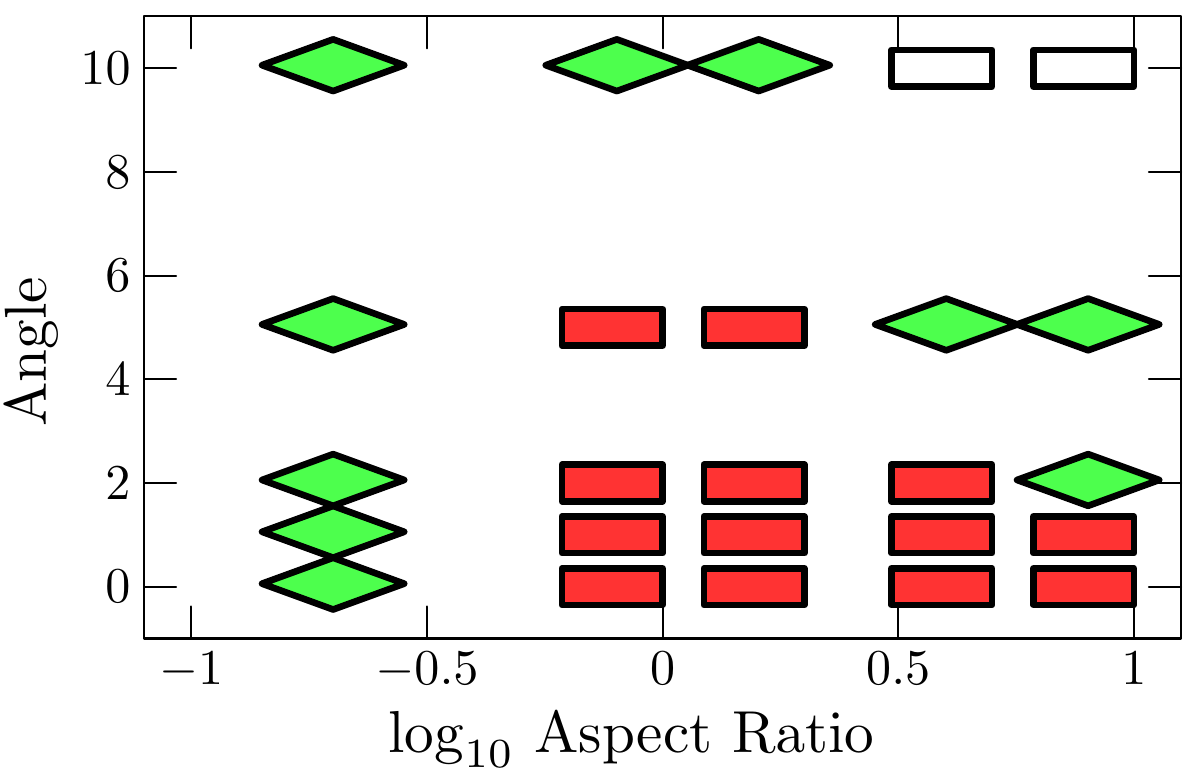}
\end{center}
\caption{The phase found in the centre of the sample at a deformation
  of $\lambda = 1.4$, for various aspect ratios and stretching angles
  relative to $\vec{n_{0}}$.}
\label{fig:microstructure_aspect_angles}
\end{figure}

The effects of aspect ratio are summarised in
Fig.~\ref{fig:microstructure_aspect_angles}, which shows the phase
present in the centre of the sample for various aspect ratios and
stretching angles. The lowest aspect ratio forms UB phase for all
stretching angles $0^{\circ}-10^{\circ}$, as the effect of the
clamps dominate the whole sample.
For higher aspect ratios the effect of the clamps on the centre of the
sample diminishes and the BB phase forms for very small
angles. However, a small deviation from stretching parallel to the
layer normal results in a reversion back to the UB phase.
Experimental studies on higher aspect ratio samples
\cite{MARC:MARC200600640} show no opacity after the stress-strain
threshold. It is tempting to associate this with a small misalignment
of the stretch axis with the layer normal, resulting in the UB or (for
large angles) the AS phase. However, the small angle X-ray scattering
does not support this as there is no reorientation of the layer normal
observed in this experiment.

Varying the thickness of the sample at a constant length-to-width
ratio results in qualitatively similar stress-strain curves and
microstructure distribution.

\section{Conclusions}

We have simulated the stretching of monodomain Sm-$A$ elastomer sheets
by using a quasi-convexified free energy model \cite{A5}.  This model
was augmented with an energy term to describe the energy of deforming
buckled layers, which is necessary to reproduce the experimentally
observed Poisson's ratios. The magnitude of this term can be measured
experimentally by a two step deformation process; first deforming the
elastomer parallel to the layer normal, then deforming perpendicular
to this direction. The modulus of the elastomer during this second
step gives the modulus of the additional energy term. 

The deformation of the elastomer in realistic, experimental geometries
was computed using finite elements. The tensile deformation of Sm-$A$
elastomer sheets of different aspect ratios, and with different angles
between the stretch axis and the layer normal were investigated. When
elongated parallel to $\vec{n_{0}}$ the majority of the sample is
predicted to form a bi-directionally buckled microstructure, except at
the clamps where uni-directional microstructure is
expected. Experimentally these microstructural differences should be
distinguishable using X-ray scattering patterns, or by examination
through a polariser-analyser pair. When elongated at a small
inclination to the layer normal the phase of the sample is sensitive
to the aspect ratio of the sample. For low aspect ratios the
bi-directionally buckled phase persists to large angles. For high
aspect ratios no buckled phase is observed in the bulk of the sample
even for small inclination angles of a few degrees between the stretch
axis and the layer normal.

\begin{acknowledgements}
  We would like to thank SEPnet for supporting this project, and Dr
  Daniel Corbett and Dr Jon Bevan for helpful discussions.
\end{acknowledgements}

\appendix

\section{Change of reference state of Smectic-$A$ model}
\label{app:oldsmarelation}

We start from the free energy density derived in Ref.~\cite{adams:05},
given by
\begin{equation}
  f_{\textrm{sm-}A} = \half \mu \textrm{Tr}\left[\lambdabold \cdot \ellbold_0 \cdot \lambdabold^T \cdot \ellbold^{-1}\right] +\half B \left( \frac{d}{d_0} -1\right)^2,
\label{eqn:smaAWfe}
\end{equation}
where $\mu$ is the shear modulus, $\lambdabold$ is the deformation
gradient \emph{starting from the smectic reference state} with
$\textrm{det}\;\lambdabold = 1$. The initial polymer conformation with
anisotropy of $r$ and mesogen alignment along the unit vector
$\vec{n}_0$ is represented by $\ellbold_0 = \Id+(r-1)
\vec{n}_0\vec{n}_0^T$. In the target state the mesogens align parallel
to $\vec{n}$, and hence the polymer conformation is described by
$\ellbold^{-1} = \Id+(\frac{1}{r}-1) \vec{n}\vec{n}^T$. $B$ is the
smectic layer modulus, $d$ is the current layer spacing and $d_0$ the
equilibrium layer spacing. The layer normal orientation denoted by the
unit vector $\vec{n}$ is assumed to deform like an embedded plane,
hence
\begin{eqnarray}
\label{eqn:appdirector}
  \vec{n} &=& \frac{\textrm{cof}\;\lambdabold \cdot \vec{n}_0}{|\textrm{cof}\;\lambdabold \cdot \vec{n}_0|}\\
  \frac{d}{d_0} &=& \frac{1}{|\textrm{cof}\;\lambdabold \cdot \vec{n}_0|},
\label{eqn:appspacing}
\end{eqnarray}
where $\vec{n}$ is the current layer normal, $\vec{n}_0$ is the
initial layer normal and $\textrm{cof} \; \lambdabold=
\lambdabold^{-T}$ denotes the cofactor of $\lambdabold$ for volume
conserving deformations. 

The free energy density in Eq.~(\ref{eqn:smaAWfe}) can be re-expressed
using the high temperature isotropic state as the reference
configuration. The deformations relative to this reference state are
given by $\ten{F}$ where
\begin{equation}
  \ten{F} = \lambdabold \cdot \ellbold_0^{1/2} r^{-1/6}.
\end{equation}
Physically we are first taking the isotropic sample in the reference
state then cooling it to the smectic state, whereupon it undergoes a
volume conserving spontaneous deformation $\ellbold_0^{1/2}
r^{-1/6}$. The deformation $\lambdabold$ is then carried out from the
smectic state. The free energy density expressed in terms of $\ten{F}$
is
\begin{eqnarray}
 f_{\textrm{sm-}A}&& =\nonumber\\
 \frac{1}{2} \mu r^{1/3}&&\left( \textrm{Tr}\;\ten{F}\cdot \ten{F}^T + kq^2\left(\frac{q}{|\textrm{cof} \;\ten{F} \cdot \vec{n}_0|} - 1\right)^2\right).
\end{eqnarray}
If we assume that $k\gg1$, then we can make the approximation
$|\textrm{cof} \;\ten{F} \cdot \vec{n}_0| \approx q$. This expression,
when converted to a dimensionless quantity by dividing by $\half \mu
r^{1/3}$ can then be approximated by Eq.~(\ref{eqn:wsmafe}).

\section{Estimation of coefficient of new term}
\label{app:modulusest}

The stiffness associated with changing the buckling wavelength of the
layers can be estimated by using a similar calculation to that of
Finkelmann \textit{et al.} \cite{finkelmann1997}.

We first calculate the free energy of a single interface between two
regions of opposite shear. Consider a Sm-$A$ film with $\vec{n}_0 =
(1,0,0)^T$. The deformation gradient tensor in the two regions is given
by
\begin{equation}
\lambdabold = \left( \begin{array}{ccc} \lambda_{xx} & 0 & \lambda_{xz}\\
0&\frac{1}{\lambda_{xx} \lambda_{zz}} & 0\\ 0 &0&\lambda_{zz}
\end{array}\right).
\end{equation}
Using Eqs.~(\ref{eqn:appdirector}) and (\ref{eqn:appspacing}) this
deformation results in the following expression for the layer spacing
and director orientation.
\begin{eqnarray}
\frac{d}{d_0} &=& \frac{\lambda_{xx} \lambda_{zz}}{\sqrt{\lambda_{xx}^2 + \lambda_{xz}^2}}
\\
\vec{n}&=&\left( \frac{\lambda_{zz}}{\sqrt{\lambda_{xx}^2 + \lambda_{xz}^2}}, 0, -\frac{\lambda_{xz}}{\sqrt{\lambda_{xx}^2 + \lambda_{xz}^2}}\right).
\end{eqnarray}
The orientation of the layer normal can be written as $\vec{n} =
\left( \cos \theta, 0, \sin \theta \right)$ where $\tan \theta =
-\lambda_{xz} / \lambda_{zz}$. If we substitute these expressions into
the Sm-$A$ free energy expression in Eq.~(\ref{eqn:smaAWfe}) we
obtain
\begin{eqnarray}
  f&& = \half \mu \left[ \lambda_{zz}^2 + \frac{1}{\lambda_{zz}^2 \lambda_{xx}^2} + \lambda_{zz}^2 \tan^2 \theta  \right.\nonumber\\
&&\left.+ (\cos^2 \theta + r \sin^2 \theta) \lambda_{xx}^2 + \frac{B}{\mu} ( \lambda_{xx} \cos \theta -1)^2\right].
\end{eqnarray}
This equation can be minimised over $\lambda_{zz}^2$, resulting in
$\lambda_{zz}^2 = \cos \theta / \lambda_{xx}$. Substituting this back
into the free energy reduces it to
\begin{eqnarray}
f &=& \half \mu \left[\frac{2}{\lambda_{xx} \cos \theta}+ \lambda_{xx}^2 (\cos^2 \theta + r \sin^2 \theta) \right.\nonumber\\
&&\left.+ \frac{B}{\mu} (\lambda_{xx} \cos \theta -1)^2\right].
\end{eqnarray}
Expanding for small $\theta$ up to quartic order, corresponding to
small rotations of the layer normal, produces the following expression
\begin{eqnarray}
f &=&\half \mu\left[p_0-p_2 \theta^2 + \third p_4 \theta^4 \right]\\
p_0 &= &\frac{2}{\lambda_{xx}}+ \lambda_{xx}^2+ \frac{B}{\mu} (\lambda_{xx}-1)^2\\
p_2 &=& -\frac{1}{\lambda_{xx}}+\lambda_{xx}^2(r-1)+\frac{B}{\mu}(\lambda_{xx}^2 - \lambda_{xx})\\
p_4 &=&\frac{1}{4} \frac{B}{\mu} \lambda_{xx} (4 \lambda_{xx} -1) + \frac{5}{4\lambda_{xx}}
+(1-r) \lambda_{xx}^2
\end{eqnarray}
In addition to the rubber elastic energy, calculation of the interface
energy requires a Frank elastic energy. For simplicity here we use the
one constant approximation, hence the total energy is
\begin{equation}
F = L_x L_y\int_0^{L_z} dz \left(\half \mu\left[p_0-p_2 \theta^2 + \third p_4 \theta^4 \right] +\half K \theta^\prime{}^2\right).
\end{equation}
It is convenient to convert distance to a dimensionless quantity using
$\xi = \sqrt{\frac{K}{\mu}}$. If we denote $t = z/\xi$, then the free
energy becomes
\begin{equation}
  F = \half \mu L_x L_y\sqrt{\frac{K}{\mu}}\int_0^{\mathcal{L}} dt \left(\left[p_0-p_2 \theta^2 + \third p_4 \theta^4 \right] +\half \dot{\theta}{}^2\right),
\end{equation}
where $L_z = {\mathcal L} \xi$. Minimization of this integral produces
the following Euler-Lagrange equation
\begin{equation}
\ddot{\theta} = -  p_2 \theta + \textstyle{\frac{2}{3}} p_4 \theta^3.
\end{equation}
Far away from the interface the director is in the energy minimum
where
\begin{equation}
\theta^2=\theta_0^2 = \frac{3 p_2}{2 p_4}.
\end{equation}
The first integral of the Euler-Lagrange equation is given by
\begin{eqnarray}
\half \dot{\theta}^2 &=& -\half p_2 \theta^2 + {\textstyle\frac{1}{6}} p_4 \theta^4 \nonumber\\
&+& \half p_2 \theta_0^2 - {\textstyle\frac{1}{6}} p_4 \theta_0^4
\end{eqnarray}
The first integral can be used to substitute for the $\dot{\theta}$
term in the free energy. If we subtract from $F$ the free energy of
the uniform state with $\theta = \theta_0$ then we obtain the free
energy of the interface
\begin{eqnarray}
  F_\textrm{int} &=& L_x L_y \sqrt{K \mu} \int_0^{\mathcal L} dt \left[ - p_2 (\theta^2-\theta_0^2) + \third p_4 (\theta^4-\theta_0^4) \right]\nonumber\\
&=& \sqrt{2 K \mu}  \; \frac{p_2^{3/2} }{p_4} L_x L_y
\end{eqnarray}

\begin{figure}[!h]
\begin{center}
  \includegraphics[width = 0.3\textwidth]{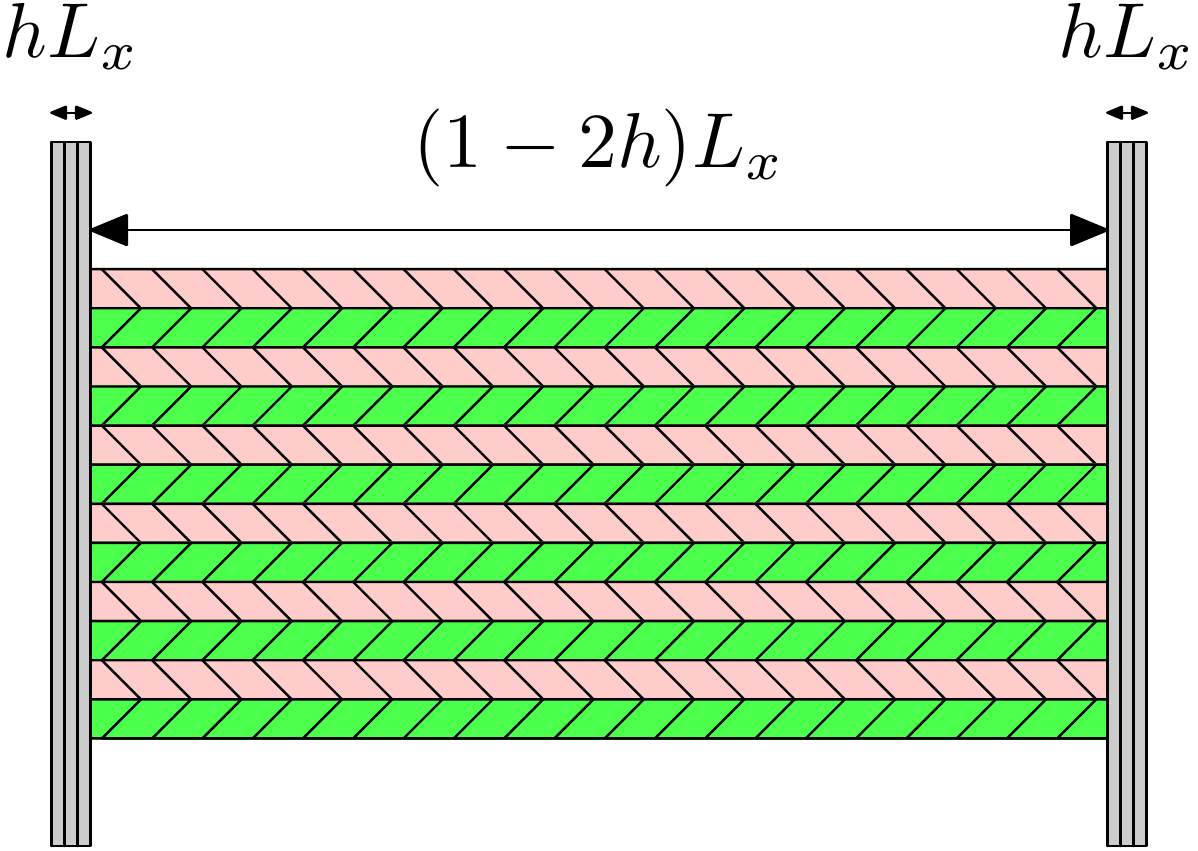}
\end{center}
\caption{To estimate the length scale of the layer buckling it is
  assumed that the sample divides into three regions as shown. The end
  regions do not contain buckled layers, whereas the central region
  does.}
\label{fig:smbucklingestimate}
\end{figure}

The wavelength of the layer buckling, and hence the stiffness of the
buckled layers can be estimated as follows. We assume that the sample
can be divided into three regions as shown in
Fig.~\ref{fig:smbucklingestimate}.
The end regions near the clamps are too constrained to buckle, so
contain layers with a fixed layer normal $(\theta = 0)$, and hence
have energy
\begin{equation}
  F_U = F_{\theta=0} = p_0(\lambda_1) V,
\end{equation}
where $\lambda_1$ is the $xx$ component of the deformation in this
region, and $V = L_xL_y L_z$ is the volume of the sample. The central
region contains smectic layers with tilt angle $\theta_0$, so has free
energy
\begin{equation}
F_R = F_{\theta= \theta_0} = p_0(\lambda_2) - \frac{3}{8} \frac{p_2(\lambda_2)^2}{p_4(\lambda_2)}
\end{equation}
where $\lambda_2$ is the $xx$ component of the deformation in this
region. If the end regions are of order $h L_x$ which in turn is
comparable to the wavelength of the layer buckling, then the number of
interfaces in the bulk is $\frac{L_z}{h L_x}$. Since the elongation of
the sample is performed by imposing a stress $\sigma$, that does work
in extending the sample, the total free energy of the system is
\begin{eqnarray}
F_T &=& (1-2 h) \mu \left(p_0(\lambda_2)- \frac{3}{8} \frac{p_2(\lambda_2)^2}{p_4(\lambda_2)} \right) V \nonumber\\
&+& 2 h \mu p_0(\lambda_1)V\nonumber\\
&-& \sigma \left( 2 h \lambda_1 + (1-2 h) \lambda_2\right) V\nonumber\\
&+& F_\textrm{int}\frac{L_z}{h L_x}.
\label{eqn:appftot}
\end{eqnarray}
If we minimise this expression over $h$, then we find the following
optimal value.
\begin{eqnarray}
h^*{}^2 &=& F_\textrm{int} \frac{L_z}{L_x V}\\
\times&&\!\!\!\!\frac{1}{\frac{3}{4} \mu \frac{p_2^2(\lambda_2)}{p_4(\lambda_2)}+ \sigma (\lambda_2 - \lambda_1)+2 \mu (p_0(\lambda_1) - p_0(\lambda_2))}\nonumber
\end{eqnarray}
To estimate the stiffness corresponding to changing the buckling
wavelength, we will assume that $h = \gamma h^*$. If we substitute
this into the Eq.~(\ref{eqn:appftot}), and calculate the second
derivative with respect to $\gamma$, then the stiffness of the sample
associated with changing the buckling wavelength is
\begin{equation}
Y = \frac{1}{2}\left. \frac{\partial^2 F_T}{\partial \gamma^2}\right| _{\gamma= 1} = \frac{F_\textrm{int}}{Vh^* L_x}
\end{equation}
If we assume that $B\gg \mu$ so that $\lambda_1 \approx 1$, then this
calculation recovers the result obtained in the text by dimensional
analysis 
\begin{equation}
Y \approx B \sqrt{\sqrt{\frac{K}{B}} \frac{1}{L_x}} f(\lambda)
\end{equation}
where $f(\lambda)$ is a function of the deformation applied.

\bibliographystyle{apsrevM}  

\ifx\mcitethebibliography\mciteundefinedmacro
\PackageError{apsrevM.bst}{mciteplus.sty has not been loaded}
{This bibstyle requires the use of the mciteplus package.}\fi

\end{document}